\documentclass[11pt]{article}
\pdfoutput=1
\usepackage{jheppub}
\usepackage[T1]{fontenc} % if needed
\usepackage{hyperref}
\usepackage{color}
\usepackage{subcaption}
\makeatletter
\def\@fpheader{\relax}
\makeatother
\newcommand{\ket}[1]{| #1 \rangle}
\newcommand{\bra}[1]{\langle #1 |}
\newcommand{\ie}{{\textit{i.e.}}~}
\newcommand{\eg}{{\textit{e.g.}}~}

\newcommand{\eq}[1]{\begin{equation}\begin{split} #1 \end{split}\end{equation}}
\newcommand{\eqs}[1]{\begin{align} #1 \end{align}}
\newcommand{\diff}[1]{{\rm d} #1 \,} % \diff{x} gives dx with correct font
\subheader{\begin{flushright}
		UTTG-08-18
\end{flushright}}

\title{Complexity-action of subregions with corners}

\author[a]{Elena Caceres,}
\author[b]{Ming-Lei Xiao}
\affiliation[a]{Theory Group, Department of Physics, University of Texas, Austin, TX 78712, USA}
\affiliation[b]{Institute of Theoretical Physics, Chinese Academy of Sciences, Beijing 100190,\\ People's Republic of China }
\emailAdd{elenac@utexas.edu}
\emailAdd{mingleix@itp.ac.cn}

\abstract{ 
In the past, the study of the divergence structure  of the holographic entanglement entropy on singular boundary regions uncovered cut-off independent coefficients. These coefficients were shown to be  universal and to encode important field theory data. Inspired by these lessons we study the UV divergences of subregion complexity-action (CA) in a region with corner (kink). We develop a systematic approach to study all the divergence structures, and we emphasize that the counter term that restores reparameterization invariance on the null boundaries plays a crucial role in simplifying the results and rendering them more transparent. We find that a general form of subregion CA contains a part dependent on the null generator normalizations and a part that is independent of them. The former includes a volume contribution as well as an area contribution. We comment on the origin of the area term as entanglement entropy, and point out that its presence constitutes a robust difference between the two prescriptions to calculate subregion complexity (-action v.s. -volume). We also find universal $\log\delta$ divergence associated with the kink feature of the subregion. Similar flat angle limit as the subregion-CV result is obtained.
}

\begin{document}
	
\maketitle

%\documentclass[10pt,aps,onecolumn,superscriptaddress,preprintnumbers,notitlepage,nofootinbib]{revtex4-1} 
%
%\usepackage{amsmath,amssymb,bbold,graphicx,natbib,bm}%,siunitx
%\usepackage{enumerate}
%\usepackage{hyperref}
%\usepackage{color}
%\usepackage{empheq}
%\usepackage{slashed}
%\usepackage[toc,page]{appendix}

%%--------------------- The Title ------------------------------------------------------------------------
%
%\title{Holographic Complexity of Subregions with Corners}
%
%\author{Elena Caceres, Ming-Lei Xiao}
%%\affiliation{Theory Group, Department of Physics, \\The University of Texas at Austin,  Austin, TX 78712 U.S.A.}
%
%\maketitle

\section{Introduction}\label{sec:intro}

%In the context of gauge/gravity duality we have learned that  quantum entanglement in a field theory encodes information about the geometry of the dual spacetime. 
The holographic principle equates entanglement entropy in a field theory to a geometrical object, called {\em holographic entanglement entropy},  in the dual spacetime \citep{Ryu:2006bv}. This relation implies that quantum entanglement encodes information about the geometry of the dual space and plays a crucial role in the program of reconstructing spacetime from boundary (field theory) data. 
However, recently we have understood that entanglement entropy cannot be the only ingredient involved in spacetime reconstruction. After a black hole is formed, the interior grows for an exponentially large time but the holographic entanglement entropy fails to reproduce this growth \cite{Maldacena:2013xja,Susskind:2013aaa}. New developments point to quantum complexity as the missing ingredient \cite{Susskind:2014rva,Susskind:2014moa}. One way to think about complexity in quantum mechanical systems is as the minimum number of ``simple'' operations needed to go from a reference state to a target state. Quantum complexity is an active area of research in quantum information but not much is known about complexity in quantum field theories. The first steps in this direction were taken in \cite{Jefferson:2017sdb, Chapman:2017rqy} where the authors investigate circuit complexity in a free scalar quantum field theory.

There are currently several proposals for the geometrical construction dual to complexity. Two of these proposals are considered more promising and have been thoroughly explored:  complexity-volume (CV) \cite{Stanford:2014jda} and complexity-action(CA) \cite{Brown:2015bva,Brown:2015lvg}. These proposals for holographic complexity aim to capture the circuit -or gate- complexity of the corresponding dual state. In the geometry side it is natural to also define a complexity not of the whole state but of a region of space, {\it i.e.} subregion complexity \cite{Alishahiha:2015rta,Ben-Ami:2016qex,Carmi:2016wjl}. 
The subregion complexity-volume proposal, subregion-CV, identifies the  subregion complexity with the maximal spatial volume bounded by the Ryu-Takayanagi surface and the boundary region. 
On the other hand, the subregion complexity-action proposal, subregion-CA, associates the complexity of a boundary region with the action evaluated on the intersection of the Wheeler-DeWitt (WDW) patch and the entanglement wedge of the given region \cite{Carmi:2016wjl}. 
Recently, in \cite{Agon:2018zso} the authors proposed several definitions for  subregion complexity in a field theory and compare their properties to the holographic proposals advanced in \cite{Alishahiha:2015rta,Ben-Ami:2016qex,Carmi:2016wjl}. They found a promising agreement of purification complexity and subregion-CA. However, the issue of which of the  holographic proposals is the correct one, or if they correspond to different definitions of complexity, is not settled yet. 

In the past, understanding the divergence structure of entanglement entropy in boundary regions with geometric singularities, {\it i.e.} regions with ``corners'',  was quite fruitful. A singular region is characterized by an opening angle $0<\Omega< \pi$. Cut-off independent coefficients, $a(\Omega)$, arising from such regions 
were studied in a variety of quantum field theories \cite{Casini:2006hu,Casini:2008as,Bueno:2015qya} (free scalars, free fermions, interacting scalars) and in holographic models \cite{Myers:2012vs,Bueno:2015rda,Bueno:2015xda,Bueno:2015lza} as well. It was found that these coefficients represent an effective measure of the degrees of freedom of the underlying CFT.  Furthermore,  it was shown that the ratio $a(\Omega\rightarrow\pi/2)/C_T$,  where $C_T$ is the central charge associated with the stress tensor $T_{\mu\nu}$, is universal for any $3D$ CFTs.

Inspired by these lessons, in this paper we study the UV divergence structure of subregion-CA of a boundary region with a kink. Our goal is to take a first step towards understanding if geometrical singularities in the boundary region also encode cutoff independent and universal contributions to subregion complexity. To calculate subregion complexity-action we have to evaluate the action in the spacetime region determined by the intersection of the entanglement wedge and the Wheeler DeWitt patch. The calculation in the case of subregions with corners is technically involved. We discuss the appropriate way to define the infrared cutoff and develop a systematic approach to calculate all the divergences. We uncover a divergence structure that is much richer than that in the subregion-CV prescription \cite{Bakhshaei:2017qud}. 
%
% We find that the leading divergence is proportional to the volume of the boundary region. Including the boundary counter terms results in several cancellations that greatly simplify the results. We identify  new  contributions coming from the singularity to the cut-off independent coefficients of  $\log$ terms. 
As in the case of subregion CA for a smooth region, there are divergences that depend on the null generator normalizations (null-norms). We find a general expression for these divergences, which includes a volume term, an entropy term, and a new term that we temptingly call the ``complexity of non-locality''. We also identify the null-norm-independent divergences: an area term and a $\log$ term. We show that the $\log$ term is the cutoff independent divergence coming from the kink feature. Our results include  the boundary counter term contributions that restore reparametrization invariance on the null boundaries. The presence of this counter term is crucial for obtaining a clean and concise final result: it cancels all $\log L$ dependences which shall not be physical for boundary theory observables; it also produces highly non-trivial cancellations of divergence structures, such as $\log^2\delta$ and $\delta^{-2}$. 
% We comment on different choices of normalizations for these null vectors. In particular, we find that taking the scale $\ell_{\alpha,\beta}$ to be associated with the size of the region, $R$, instead of the UV cutoff, $\delta$, leads to a desirable complexity behavior.  We also point out a relation similar to the Bousso bound that is obeyed in the case of a kink singularity. 

Our detailed results can serve as a benchmark for proposals of subregion complexity in field theory. 
Furthermore, the systematic approach to study the divergence structures that we develop here can be easily extended to higher dimensions and more general geometric singularities. 

This paper is organized as follows. In Section~\ref{sec:EE} we review the relevant ideas of entanglement entropy in subregions with corners and their significance. In Section~\ref{sec:subC} we review  the definitions of subregion-CV and subregion-CA, and the subregion-CV result for a kink region \citep{Bakhshaei:2017qud}. Sections~\ref{sec:CA_cpt} and \ref{sec:result} constitute the main parts of this paper. In Section~\ref{sec:CA_cpt} we setup the problem, point out some subtleties and outline the steps of the calculation. In Section~\ref{sec:result} we present the final result and discuss its various properties. Section~\ref{sec:conclusions} contains the conclusions and future directions. All the technical details of the calculations are presented in two appendices.

\section{Subregions with geometric singularities}
\label{sec:EE}

Spatial subregions in the boundary theory that contain geometric singularities are known to have interesting contributions to the entanglement entropy. In quantum field theory, the entanglement entropy has an area law behavior. But the coefficients of the leading order area law contribution depend on the UV regularization of the theory. On the other hand, there are subleading contributions that are independent of the UV regularization and thus,  contain unambiguous information about the boundary theory \cite{Solodukhin:1994yz}. These contributions were later shown to be universal for a large class of CFTs \cite{Casini:2006hu}. When the boundary has sharp features or singularities it was found in \cite{Bueno:2015rda,Bueno:2015xda} that there are additional contributions that are cutoff independent and universal . In this section we will review some results related to entanglement entropy in regions with corners.

The metric of $d+1$ dimensional  AdS space in Poincare patch is,
\eq{\label{eq:metric_polar}
	ds^2 = \frac{L^2}{z^2}\left[-\diff t^2 + \diff z^2 +\diff\rho^2 + \rho^2(\diff\theta^2 + \sin^2\theta\diff\Omega_n^2)\right].
}
where $n=d-3$. A cone is an example of a singular region on the boundary. In general, cones in different dimensions can be parametrized as,
\eq{\label{eq:cone_range}
	c_n = \{ t=0,\ |\theta|\leq\Omega,\ 0\leq\rho < \rho_{\rm IR} \}.
}
where  $\rho_{\rm IR}$ is an IR cutoff that has to be taken to infinity at the end of the calculation. The cone  $c_n$ has a scaling symmetry along the radial direction. Due to this symmetry, the Ryu-Takayanagi (RT) surface\footnote{More precisely, one should take the Hubeny-Rangamani-Takayanagi (HRT) surface \citep{Hubeny:2007xt} to compute entanglement entropy, which reduces to the RT surface in a time translational symmetric setup.}\cite{Ryu:2006bv} should take the form
\eq{
	z_{RT} = \rho\,  h(\theta). \label{eq:h_RT}
}
The function $h(\theta)$ characterize the shape of the RT surface, which has a maximum $h(0) \equiv h_0$, and vanishes at the boundary $h(\Omega)=0$. The entanglement entropy of $c_n$ is given by the area functional
\eq{\label{eq:RT_area}
	S_n = L^{n+2}\Omega_{n}\int\frac{\diff\rho}{\rho}\ \int\diff\theta\ \sin^{n}\theta\frac{\sqrt{1+h^2+h^{\prime 2}}}{h^{n+2}} ,
}
whose extremality condition determines the shape function $h(\theta)$. Since $h(\theta)$ specifies the RT surface \eqref{eq:h_RT}, this function also plays an important role in the complexity calculations of the following sections, hence we derive its property below for future convenience. 

In this paper we focus on $d=3$ , {\it i.e.}  $n=0$. The $c_0$ cone is also referred to as  a kink. In this case the integrand in \eqref{eq:RT_area} is independent of $\theta$ and the area functional has an integration constant
\eq{\label{eq:K}
	K = \frac{1+h^2}{h^2\sqrt{1+h^2+h^{\prime 2}}} = \frac{\sqrt{1+h_0^2}}{h_0^2}.
}
%In terms of $K$, one can express derivatives of $h(\theta)$ as functions of $h$, hence
%\eq{\label{eq:solh}
%	h'(\theta) &= \frac{\sqrt{(1+h^2)(1+h^2-K^2h^4)}}{Kh^2}, \\
%	h''(\theta) &= -\frac{2+2h^2+K^2h^6}{K^2h^5}.
%}
The opening (half) angle $\Omega$ can then be written as a function of $h_0$, 
\eq{\label{eq:def_omega}
	\Omega(h_0) = \int_0^{h_0}\frac{\diff h}{h'(\theta)} = \int_0^{h_0} \frac{Kh^2}{\sqrt{(1+h^2)(1+h^2-K^2h^4)}}\diff h \equiv \int_0^{h_0} \omega(h) \diff h.
}
The inverse function $h_0(\Omega)$ does not have a simple form, thus we treat $h_0$ as the angle variable throughout the calculations. The asymptotic behavior at small or flat angle limits are important, which we show explicitly here
\eq{\label{eq:open_angle}
	&\Omega(h_0\to0) = \gamma h_0 + \mathcal{O}(h_0^3), \\
	&\Omega(h_0\to\infty) = \frac{\pi}2( 1-h_0^{-1} ) + \mathcal{O}(h_0^{-2}),
}
where
\eq{\label{eq:def_gamma}
	\gamma = \frac{\sqrt{\pi}\Gamma(3/4)}{\Gamma(1/4)} \approx 0.599.
}
The flat angle case, $\Omega = \pi/2$, has no singular kink feature, hence any ``kink contributions'' that we identify should vanish in this limit. 

We should also distinguish the case with $\Omega < \pi/2$ and that with $\Omega > \pi/2$, the former being a convex kink and the latter a concave kink. Since the RT surface is the same for both cases, one does not need to worry about it while dealing with entanglement entropy; but the complexity computation involving entanglement wedge depends on which side of the RT surface is identified as inside.

% An natural extension of the cones are cones with creases, denoted by $c_n\times\mathbb{R}^m$. The simplest example is $k\times\mathbb{R}$ . 

In \cite{Bueno:2015lza} the authors found that cone regions contribute to universal terms in the entanglement entropy. These contributions introduce new $\log$ or $\log^2$ terms that are cutoff independent,
\eq{\label{eq:uni_entropy}
	S^{\rm univ}(V) = \left\{
	\begin{array}{ll} 
		(-1)^{\frac{d-1}2}a^{(d)}(\Omega)\log(R/\delta),\quad 		& d\ {\rm odd}, \\
		(-1)^{\frac{d-1}2}a^{(d)}(\Omega)\log^2(R/\delta),\quad	 	& d\ {\rm even}.
	\end{array}
	\right.
}
The functions $a^{(d)}(\Omega)$ are functions of the opening angle $\Omega$. Since we are dealing with a pure state,   $a^{(d)}(\Omega)=a^{(d)}(\pi -\Omega)$.  An additional restriction on $a^d(\Omega)$ comes from the fact that when $\Omega\rightarrow \pi/2$ we are in the smooth limit, with no singularity, and therefore $ a^{(d)}(\Omega=\pi/2)=0$. These constraints imply that in the large anlge limit $a^d(\Omega)$ is of the form
\eq{\label{eq:uni_entropy_coeff}
	a^{(d)}(\Omega\to\pi/2) = \sigma^{(d)}\left(\pi-2\Omega\right)^2.
}
Thus, the conical singularity  introduces  a set of coefficients $\sigma^{(d)}$ that encode cutoff-independent information about the CFT.   Remarkably, this same behavior for  $\sigma^{(d)}$ was found for field theory calculations of entanglement in regions with sharp corners \citep{Bueno:2015qya}.  Furthermmore, holographically, it can be shown that  $\sigma^{(d)}$ is purely determined by the boundary stress tensor charge $C_T$. As mentioned in the Introduction, the motivation of the present work is to understand if similar cutoff independent and possibly universal contributions are present in the case of subregion complexity-action.

%--------------------- Sec 3 ------------------------------------------------------------------------

\section{Subregion complexity }
\label{sec:subC}
%\subsection{Extension of CA duality conjecture}

Currently there are two proposals for holographic subregion complexity: subregion-CV and subregion-CA. Two basic criteria are met by both of these proposals: 1) They recover the original holographic state complexity in the limit when the region is the whole boundary space; 2) Since a boundary subregion state should be holographically dual to its entanglement wedge \cite{Headrick:2014cta}, the volume in subregion-CV or the action in subregion-CA should both be evaluated within this bulk region to reflect this correspondence. 

In the CV approach, one takes the maximal spatial volume bounded by the boundary subregion and its HRT surface, which is of course contained in the entanglement wedge. Also, if we take the subregion to be the whole boundary we clearly recover the original CV-complexity \cite{Stanford:2014jda}.  Subregion-CV complexity was investigated in \citep{Alishahiha:2015rta} for smooth subregions and in  \cite{Bakhshaei:2017qud} for subregions with corners. In particular, for a 3 dimensional kink the subregion-CV complexity is,
\eq{\label{eq:cv_kink}
	\mathcal{C}_k = \frac{L^2}{8\pi G} \left[\frac{\Omega}{2}\frac{R^2}{\delta^2} - \alpha(h_0)\log\frac{R}{\delta}\right].
}
Interestingly, besides the regular volume contribution, a new term with log divergence appears as the kink contribution. The coefficient
\eq{
	\alpha(h_0) = \int_0^\Omega\frac{\diff\theta}{h^2} = \int_0^{h_0}\frac{\omega(h)}{h^2}\diff h
}
has the limiting behavior
\eq{\label{eq:CV_result}
&\lim_{h_0\to0}\alpha(h_0) = \frac{\gamma'}{h_0}, \quad \gamma' = \int_0^1\frac{2}{\sqrt{1-x^4}}\diff x \approx 2.622, \\
&\lim_{h_0\to\infty}\alpha(h_0) = \frac{\pi}{h_0}. 
}
Note that there is no $\delta^{-1}$ divergence (or area term) in this setup. It is shown \cite{Carmi:2016wjl} that for time-symmetric configuration, subregion-CV cannot have an area term. 

In \cite{Carmi:2016wjl} the authors proposed that the subregion-CA complexity is given by the action evaluated on the intersection of the entanglement wedge of the subregion and the WdW patch of the boundary time slice.
\begin{figure}[!ht]
	\centering
	\includegraphics[height=3in]{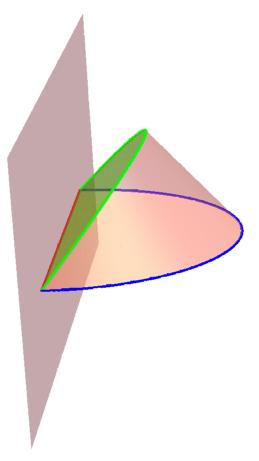}
	\caption{The intersection of the entanglement wedge and WDW patch. For clarity, we only show  the $t>0$ part of the region, $\mathcal{V}_+$. This is the  upper half of the full region $\mathcal{V}$. The lower half $\mathcal{V}_-$ is symmetric with $t\rightarrow -t$. In the $\delta \rightarrow 0$ limit, the red line represents both, the boundary region $\mathcal{A}$ and the surface $W$ on the cutoff surface. The blue curve represents the HRT surface $E$. The green and pink surfaces represent the null hypersurfaces $\mathcal{W}^+$ and $\mathcal{E}^+$, and their intersection at the green curve is the surface $J^+$. }\label{fig:calV}
\end{figure}

Fig~\ref{fig:calV} schematically shows the upper half of the relevant bulk region. The spatial region is represented by a red line, the HRT surface by a blue curve and the light sheet associated to it, {\it i.e.} the  boundary of the entanglement wedge, is the pink surface denoted by $\mathcal{E}$. The green surface represents the light sheet associated to the boundary interval and is, therefore, the boundary of the WdW patch $\mathcal{W}$. In addition to the region depicted in Fig~\ref{fig:calV} the full region of interest contains a symmetric $(t\rightarrow-t)$ lower half.

% We will show in Section 4) that the subregion-CA complexity of a region with a kink presents a much richer divergence structure than \ref{eq:cv_kink}. Also, new cut-off independent contributions, that are expected to encode physical information, appear.

%--------------------- Sec 4 ------------------------------------------------------------------------

\section{Complexity-action of a region with 3d kink}
\label{sec:CA_cpt}

\subsection{Setup}
\label{sec:setup}

To compute the holographic complexity of a 3d kink region $\mathcal{A}$ in subregion-CA approach, one has to first define the bulk region $\mathcal{V}$ on which we compute the action. As described in the previous section, this region has a boundary that consists of 4 hypersurfaces
\eqs{
	& \mathcal{W}^\pm:\ z = \delta \pm t,\ (x,y)\in \mathcal{A}; \label{eq:def_W} \\
	& \mathcal{E}^\pm:\ X_{\pm}^{\mu}(\lambda) = X_0^{\mu} + f(\lambda)V_{\pm}^{\mu}. \label{eq:para_E}
}
with $\pm$ signs labelling the upper half $t>0$ and lower half $t<0$. 
The $\mathcal{W}^\pm$ shall be defined more carefully in Sec.(\ref{sec:ir}) due to the IR cutoff subtlety.
% For calculational purposes it is more convenient to define  $\mathcal{W}^\pm$ in a parametric way similar to  $\mathcal{E}^\pm$ in \eqref{eq:para_E}. We will do that in section \ref{sec:ir}. 

In our notation $X_{\pm}^\mu$ are Cartesian coordinates of points on $\mathcal{E}^{\pm}$ generated by inward normal light rays from the HRT entangling surface $E = \mathcal{E}^+ \cap \mathcal{E}^-$. Points on $E$ are given by \eqref{eq:h_RT}, 
\eq{
	X_0^\pm = \rho h(\theta)\mathbf{\hat{z}} + \rho\cos\theta\mathbf{\hat{x}} + \rho\sin\theta\mathbf{\hat{y}}.
	% X_0^t = 0, \quad X_0^z = \rho h(\theta), \quad X_0^x = \rho\cos\theta, \quad X_0^y = \rho\sin\theta.
}
The inward null normal vectors $V_{\pm}^\mu$ that generate $\mathcal{E}^{\pm}$ thus satisfy 
\eq{
	V_{\mu}V^{\mu} = 0 ,\quad V_{\mu}\frac{\diff X^{\mu}}{\diff \rho} = 0 ,\quad V_{\mu}\frac{\diff X^{\mu}}{\diff \theta} = 0.
}
These are solved by:
\eq{
	& V^<_{\pm} = \pm\sqrt{1+h^2+h^{\prime 2}}\mathbf{\hat{t}} - \mathbf{\hat{z}} + (h\cos\theta - h'\sin\theta)\mathbf{\hat{x}} + (h\sin\theta + h'\cos\theta)\mathbf{\hat{y}} \\
	& V^>_{\pm} = \pm\sqrt{1+h^2+h^{\prime 2}}\mathbf{\hat{t}} + \mathbf{\hat{z}} - (h\cos\theta - h'\sin\theta)\mathbf{\hat{x}} - (h\sin\theta + h'\cos\theta)\mathbf{\hat{y}}
}
The two solutions represent the different orientations of the kink, as a kink and its complement share the same HRT surface. For convex kinks with $\Omega < \pi/2$, one takes $V^<$; for concave kinks with $\Omega > \pi/2$, one takes $V^>$.

Due to the conformal flatness in Poincare patch, the light rays are straight lines, with linear coefficients $f(\lambda)$. This reparameterization function $f$ is determined by requiring $\lambda$ to be an affine parameter, so that $\diff X^{\mu}(\lambda) / \diff\lambda$ constitute a geodesic congruence. This condition is solved by
\eq{\label{eq:sol_f}
	f(\lambda) = \beta\frac{\lambda\rho^2h^2}{L^2\pm \lambda\rho h}
}
where $\pm$ are for convex and concave kinks respectively. $\beta$ is a constant not fixed by the geodesic equation for $\diff X^{\mu}(\lambda) / \diff\lambda$.

We further define $W = \mathcal{W}^+ \cap \mathcal{W}^-$ and $J^{\pm} = \mathcal{E}^\pm \cap \mathcal{W}^\pm$. $W$ is on the cutoff surface $z=\delta$ which approaches $\mathcal{A}$ in the $\delta \to 0$ limit. These geometrical objects can be visually seen in Fig.~\ref{fig:calV}. Other relevant geometrical computations on them are presented in Appendix~\ref{ap:geo}.

\subsection{Subtleties in the complexity-action calculation}
\label{sec:issues}

Before we plunge into the calculation of  the action in $\mathcal{V}$, we want to point out a few issues that are worth careful considerations. First, the null hypersurface, $\mathcal{E}$, entering the definition of $\mathcal{V}$ can present caustics. We show that for the kink singularity we consider here, the caustics are outside the region of interest and therefore pose no problem.  Second, since the subregion we consider is not closed we need a IR cutoff. We show that a consistent IR cutoff should be carefully chosen for subregion-CA computations. We then introduce the important ingredients of action computation for bulk regions with boundaries, especially with null boundaries. In particular, joint contributions and counter terms for reparameterization invariance are defined.

\subsubsection{Caustics}
\label{sec:caustic}

According to the focusing theorem, lightsheets end on caustics in finite amount of affine time. If the lightsheet $\mathcal{E}^\pm$ end before they intersect with $\mathcal{W}^\pm$, the caustics would be part of the bulk region and we need to take special care of them.

The expansion rate $\Theta$ of the lightsheet congruence is 
\eq{\label{eq:def_expansion}
	\Theta_\pm(\lambda) = \frac{1}{\sqrt{g_\pm}}\frac{\partial\sqrt{g_\pm}}{\partial\lambda}
}
where $g_{\pm,\alpha\beta}(\lambda)$ are the induced metric on equal $\lambda$ slice of the light sheet and $\lambda$ is the affine parameter of the geodesic $X_\pm^{\mu}(\lambda)$. We solve the geodesic equation to obtain $X_\pm^{\mu}(\lambda)$, determine the induced metric and obtain $\Theta_\pm$\footnote{See Appendix \ref{ap:geo}) for details.} %$k^{\mu}\nabla_{\mu}k^{\nu}=0$ for $k^{\mu}\equiv\partial X_\pm^{\mu} / \partial\lambda$, which yields
%\eq{
%	f^<(\lambda) = \frac{\lambda\rho^2h(\theta)^2}{L^2+\lambda\rho h(\theta)}, \qquad f^>(\lambda) = \frac{\lambda\rho^2h(\theta)^2}{L^2-\lambda\rho h(\theta)}.
%}
%One can then solve for the expansion
\eq{\label{eq:expansion}
	\Theta_\pm = -\frac{2\lambda\rho^2h^2}{L^4-\lambda^2\rho^2h^2}.
}
Caustics occur when $\Theta_\pm$ diverges, which is at 
\eq{
	\lambda = \lambda_c \equiv \frac{L^2}{\rho h}.
}
Note that this result does not depend on orientation of the kink. In order to see if we encounter caustics  before reaching the intersection $J^\pm$, we solve for the $\lambda$  on $J^\pm$ by combining \eqref{eq:def_W} and \eqref{eq:para_E},
 \eq{ X_\pm^z(\lambda_*) = \delta \pm X_\pm^t(\lambda_*)}
 where,
\eq{\label{eq:lambdaE_star}
	\lambda^<_* = \frac{\lambda_c(\rho h -\delta)}{\rho h\sqrt{1+h^2+h^{\prime 2}} + \delta}, \qquad \lambda^<_* = \frac{\lambda_c(\rho h -\delta)}{\rho h\sqrt{1+h^2+h^{\prime 2}} - \delta}.
}
It is easy to see that $\lambda_* < \lambda_c$ for any $\rho$ and $\theta$, and for both convex and concave kinks. Therefore, the caustics are always outside  the region $\mathcal{V}$ and we don't need to worry about them.

\subsubsection{IR cutoff}
\label{sec:ir}

\begin{figure}[!ht]
	\centering
	\includegraphics[width=3in]{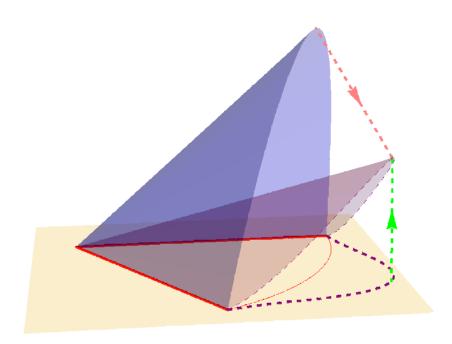}
	\caption{ Region $\mathcal{V}$ with both spatial directions shown explicitly -- the time direction is suppressed. The yellow plane is the boundary space, with the red curve denoting  the kink. The blue surface is the HRT surface $E$ and the purple one is $J$. The null hypersurfaces $\mathcal{E}$ and $\mathcal{W}$ are the volume between $E$ and $J$ and the volume between the boundary and $J$, while the red and green dashed arrows are typical null rays on them. It is clear that if we choose $\rho = R$ constant as IR cutoff on $E$ as shown here, the red null ray matches a corresponding green null ray only when the IR cutoff on the boundary is given by  the purple dashed curve instead of the naive orange circle.}\label{fig:bulk_closeness}
\end{figure}

Another tricky issue is how to choose the IR cutoff of the bulk region. It is tempting to naively use the constant cutoff $R$ for the radius coordinate on both $\mathcal{A}$ and $E$. However, with this choice of IR cutoff the hypersurfaces $\mathcal{E}^{\pm}$ and $\mathcal{W}^{\pm}$ do not match exactly to enclose the region. Thus, we can only set a constant cutoff $R$ for either $\mathcal{A}$ or $E$, while the cutoff on the other hypersurface is determined by the closeness of $\mathcal{V}$. This is illustrated by the dashed arrows in Fig.\ref{fig:bulk_closeness}. Specifically, we find the other IR cutoff by following the null rays on $\mathcal{E}$ and $\mathcal{W}$ originating from the constant cutoff. As we have already parameterized $\mathcal{E}^{\pm}$ in terms of coordinates $w_\alpha = (\rho,\theta)$ on $E$, it is more convenient to work with a constant IR cutoff $\rho = R$ on $E$, but $\rho$ should no longer be understood as the usual projected radial coordinate as in $(t,z,\rho,\theta)$ coordinate system, but rather some new induced coordinate in this parameterization\footnote{
We can also choose a non-trivial IR cutoff $\rho_{\rm IR} = R(\theta)$ that varies with $\theta$. For instance, by appropriate choice, a certain $R(\theta)$ would induce a constant IR cutoff on $\mathcal{A}$, resulting in a nice sector-shaped subregion with expected volume $R^2\Omega$. However, it increases the complexity of the computation, and does not affect the cutoff independent contributions from the kink that we are interested in. 
% Thus we insist on using a more convenient cutoff $\rho = R$ for our computation, at the cost of having an unfamiliar volume formula to appear later.
}. 

We want to use coordintes $w_\alpha = (\rho,\theta)$ to parameterize the whole region $\mathcal{V}$. 
% so that the parameter range bounded by constant cutoff $R$ sets the true range of $\mathcal{V}$.
  So far we have used a parametrization along $\mathcal{E}$ up to the surface $J$ as $X^{\mu}(\lambda)$, $0<\lambda<\lambda_*$. We continue along $\mathcal{W}$ towards the boundary surface $W$ by following the null rays that generate $\mathcal{W}$,
\eq{
	U_{\pm} = \mp\mathbf{\hat{t}} - \mathbf{\hat{z}}, 
}
with integral curve
\eq{\label{eq:para_W}
	\mathcal{W}^{\pm}: \tilde{X}_{\pm}^{\mu}(\eta) = X_{\pm}^{\mu}(\lambda_*) + \eta U_{\pm}^{\mu}.
}
It is an equivalent, but computational more practical way of defining $\mathcal{W}^\pm$ than \eqref{eq:def_W}.
The starting point $X_{\pm}^{\mu}(\lambda_*)$ is at the joint surface $J^{\pm}$, while the ending at $W$ is solved via $\tilde{X}_{\pm}^z(\eta_*)=\delta$ as,
\eq{
	\eta^<_* = \frac{1+h^2}{H_+}(\rho h - \delta), \qquad \eta^>_* = \frac{1+h^2}{H_-}(\rho h - \delta),
}
where
\eq{
	H_\pm \equiv 1+h^2\pm Kh^2\label{eq:Hpm}
}
is a shorthand notation used throughout the paper. The surface $W$ is parameterized as $\tilde{X}^\mu(\eta_*, \rho,\theta)$ where $\rho < R$ sets the cutoff boundary, as shown in Fig.~\ref{fig:bulk_closeness} by the purple dashed curve.

%
%\begin{figure}[!ht]
%	\centering
%	\includegraphics[width=0.45\textwidth]{plots/pIR1}
%	\\
%	\includegraphics[width=0.45\textwidth]{plots/pIR3}
%	\caption{We plot the shape of surface $W$ for different openning angles, labeled by corresponding values of $h_0$. The cutoff $R$ is taken to be 10. The yellow shade represent the projection of $E$ which has the shape of a sector with radius $R=10$, and the blue shade is the true range of $W$ with the same IR cutoff.}\label{fig:IRregion}
%\end{figure}
%
%In Fig.~\ref{fig:IRregion}, we show exact shape of $W$ for different kink opening angles. The blue shades are the true range of $W$ while the yellow shades are the projection of $E$ on the boundary that has constant radial cutoff.
%
% When the opening angle is small, the extra piece is small and 
% the volume is dominated by the sector shape of the projection; for large opening angle, the shape becomes much larger than the projection. In the limit of flat angle kink, $W$ will have the shape of a half strip. For any opening angles, the two regions share the same boundary surface as the two straight sides. 

Now that we have parameterized all the boundary hypersurfaces by $(\rho,\theta)$, we can naturally extend the parameterization to the whole region $\mathcal{V}$ together with $\zeta=f(\lambda)$ and $\eta$:
\eq{\label{eq:new_cor}
	\mathcal{V}_{\pm}^{\mu}(\zeta,\eta,\rho,\theta) = X_0^{\mu}(\rho,\theta) + \zeta V_{\pm}^\mu(\rho,\theta) + \eta U_{\pm}^\mu
}
with range of parameters
\eq{\label{eq:range}
	& 0 < \rho < R,\quad 0 < h < h_0,\quad \rho h > \delta, \\
	& 0 < \zeta < \zeta_* \equiv f(\lambda_*),\quad 0 < \eta < \eta_*,\quad \frac{\eta}{\eta_*} < \frac{\zeta}{\zeta_*}.
}
The last condition is to restrict the parameterization $\mathcal{V}_\pm$ to $\pm t > 0$ regime, and hence $\mathcal{V} = \mathcal{V}_+ \sqcup \mathcal{V}_-$. Due to time reflection symmetry, the action should be twice the action in either of $\mathcal{V}_\pm$.

For convenience, we also write the bulk reparameterization explicitly in the form of a coordinate transformation (for convex kink):
\eq{\label{eq:cor_trans}
	& t = \pm\zeta\sqrt{1+h^2+h^{\prime 2}} \mp \eta, \\
	& z = \rho h - \zeta - \eta, \\
	& x = \rho\cos\theta + \zeta(h\cos\theta - h'\sin\theta), \\
	& y = \rho\sin\theta + \zeta(h\sin\theta + h'\cos\theta), \\
	& \text{metric\ for\ coordinate\ system\ (t,z,x,y):}\quad g^0_{\mu\nu} = \frac{L^2}{z^2}\eta_{\mu\nu},
}
with which one can easily derive the bulk metric in this coordinate system.

\subsubsection{Higher codimension manifolds on the boundary}
\label{sec:highCoD}

We can write down the most general form of gravitational action as
\eq{\label{eq:I_general}
	I_{\rm grav} = \frac{1}{8\pi G}\sum_{d=0}^D\sum_{i}^{n_d}\int_{\Sigma_i^{(d)}} \sqrt{\sigma_i^{(d)}} \phi_i^{(d)},
}
where $\Sigma_i^{(d)}$ are dimension $d$ manifolds with metric $\sigma_i^{(d)}$ that are relevant for the spacetime region in which we compute action. In particular, $\Sigma_i^{(D)}$ are the bulk regions with $D$ the total dimension of spacetime. For this bulk integration, the integrand is, say, the Einstein-Hilbert term, hence $\phi^{(D)} = \frac{1}{2}R - \Lambda $. When $d<D$, $\Sigma_i^{(d)}$ are manifolds on the boundaries of $\Sigma_i^{(D)}$. For example, a cube is a region $\Sigma^{(D=3)}$ with boundary manifolds $\Sigma^{(2)}_{1,\cdots,6}$ as the surfaces, $\Sigma^{(1)}_{1,\cdots,12}$ as the edges and $\Sigma^{(0)}_{1,\cdots,8}$ as the vertices. For $\Sigma^{(D-1)}$, as long as it is non-null, we have the usual York-Gibbons-Hawking (YGH) term $\phi^{(D-1)} = K$, \emph{i.e.} the trace of the extrinsic curvature.

Note that for Lorentzian spacetime there can be null manifolds on the boundary, which has degenerate metric. Contributions from null $\Sigma_i^{(D-1)}$ has been studied \citep{Parattu:2015gga}, which shows that the volume form should be modified and the integrand should be the surface gravity:
\eq{
	\label{eq:kappa}
	\frac{1}{8\pi G}\int_{{\rm null}\ \Sigma_i^{(D-1)}}\diff\lambda \diff x^{D-2}\sqrt{\sigma_i^{(D-2)}} \kappa,
}
where $\lambda$ is the null parameter and $\kappa$ is the surface gravity associated with null vector field $\partial/\partial\lambda$. Inspired by the CA conjecture, further studies have been done for contributions from higher codimension manifolds. The codimension 2 non-null manifolds, called joints, were studied in \citep{Lehner:2016vdi} which provides the integrand as
\eq{
	\label{eq:joint}
	\phi_{\Sigma_i^{(D-1)} \cap \Sigma_j^{(D-1)}}^{(D-2)} \equiv a_{ij} = \pm\log\frac{k_i \cdot k_j}{2},
}
where the joint is specified as intersection of two codimension 1 manifolds $\Sigma_{i,j}^{(D-1)}$. We only present here the case when both $\Sigma_{i,j}^{(D-1)}$ are null, as it is the only relevant case for our computation. The vectors $k_{i,j} = \partial/\partial\lambda_{i,j}$ are null generators of the codimension 1 manifolds, and the sign depends on the orientation of the intersection.

Note that, in principle,  higher co-dimension singular can also be present. It was argued in \citep{Chapman:2016hwi} that a high codimension conical singularity can be regulated to a geometry with only lower dimensional singular surface, and through the regulation it was shown that the conical singularity does not contribute to the action. However,  there is no staightforward way to generalize the regulation method  to deal with general singular features, like polyhedral singularity (the intersection of several hypersurfaces). Polyhedral singularity naturally appear in any subregion CA computation, the common one being the codimension 3 manifold $\partial W$ sitting at the intersection of all of the four $\mathcal{E}^\pm$ and $\mathcal{W}^\pm$ hypersurfaces. In our case when there is additional singular feature on the surface of the boundary subregion, even higher codimension singularities are present on  $\partial W$. Understanding if polyhedral singularities can be reduced via some type of regularization and what excatly their contribution is is an issue that deserves further study.

\subsubsection{Reparameterization invariance null hypersurfaces}\label{sec:ct}

There is a problem of reparameterization in the formula \eqref{eq:joint}: the action depends on the unphysical reparameterization of the null direction $\lambda \to f(\lambda)$. 
It is pointed out in \citep{Lehner:2016vdi} that the following choice of counter term could remove the effects of the reparameterization
\eq{\label{eq:ct}
	\Delta I = -\frac{1}{8\pi G}\int_{{\rm null}\ \Sigma_i^{(D-1)}}\diff\lambda \diff x^{D-2}\sqrt{\sigma_i^{(D-2)}} \left[ \Theta \log\frac{cL|\Theta|}{2} \right] ,
}
where $\Theta$ is the expansion rate on the null hypersurface, defined previously. Here I have chosen dimensionless combination $cL|\Theta|$ as the argument of $\log$, with $c$ an arbitrary constant. Different choices of $c$ indicate different renormalization conditions. The explicit $L$ dependence here guarantees the cancellation of $\log L$ dependence in the final result. 
To see that, we write the integrand as
$$ \Theta \log\frac{cL|\Theta|}{2} = \Theta \log\frac{\delta|\Theta|}{2} + \Theta \log\frac{cL}{\delta}, $$
where the first term contributes a non-ambiguous term that makes the action reparameterization invariant, and the second term integrates to a volume difference:
\eq{ \label{eq:renorm}
	& -\frac{1}{8\pi G}\int_{{\rm null}\ \Sigma_i^{(D-1)}}\diff\lambda \diff x^{D-2}\sqrt{\sigma_i^{(D-2)}(\lambda)} \left[ \Theta \log\frac{cL}{\delta} \right]\\
	= & -\frac{1}{8\pi G}\log\frac{cL}{\delta}\left.\int_{\Sigma_i^{(D-2)}(\lambda)} \diff x^{D-2}\sqrt{\sigma_i^{(D-2)}(\lambda)}\right|_{\lambda_{\rm min}}^{\lambda_{\rm max}} \\
	= & -\frac{1}{8\pi G}\log\frac{cL}{\delta}\left(V_{\Sigma^{(D-2)}_{i1}} - V_{\Sigma^{(D-2)}_{i2}} \right)
}
where the second line used the definition eq~\eqref{eq:def_expansion}.
$\Sigma^{(D-2)}_{i1,2}$ are the codim-2 manifolds at the two ends of the null parameter $\lambda$, hence they are joint surfaces with other codim-1 boundary surfaces. Other $\log L$ dependences come exactly from these joint contributions. 
Suppose we define the generators of $\Sigma^{(D-1)}_i$ as $k_i = \alpha_i \bar{k}_i$ where $\bar{k}_i$ are normalized with some particular rule, \eg~the rule used in \citep{Lehner:2016vdi} that it has unit inner product with time-like normal vector at the boundary. 
We leave a constant $\alpha_i$ unfixed to indicate the arbitrariness of the rule, which will be soon proved to be reparameterization invariant with the addition of the counter term eq~\eqref{eq:ct}. The relevant joint contributions for null surface $\Sigma^{(D-1)}_i$ are thus
\eq{
	& I_{\Sigma^{(D-2)}_{i1}} + I_{\Sigma^{(D-2)}_{i2}} \\
	= & \frac{1}{8\pi G}\int_{\Sigma^{(D-2)}_{i1}}\diff x^{D-2}\sqrt{\sigma^{(D-2)}(\lambda_{\rm max})} \left[ \log\frac{\alpha_i L}{\delta} + \log\frac{\alpha_1 L}{\delta} + \log\frac{\bar{k}_i^{\mu}g_{\mu\nu}\bar{k}_1^{\nu}\delta^2}{2L^2} \right] \\
	& -\frac{1}{8\pi G}\int_{\Sigma^{(D-2)}_{i2}}\diff x^{D-2}\sqrt{\sigma^{(D-2)}(\lambda_{\rm min})} \left[ \log\frac{\alpha_i L}{\delta} + \log\frac{\alpha_2 L}{\delta} + \log\frac{\bar{k}_i^{\mu}g_{\mu\nu}\bar{k}_2^{\nu}\delta^2}{2L^2} \right] \\
	= & \frac{1}{8\pi G}\log\frac{\alpha_i L}{\delta}\left(V_{\Sigma^{(D-2)}_{i1}} - V_{\Sigma^{(D-2)}_{i2}} \right) + \cdots
}
where $\cdots$ denote contributions from the second and third terms in the square brackets. The second term 
% exists only when $\Sigma^{(D-2)}_{ij}$ is the intersection with another \textbf{null} hypersurface $\Sigma^{(D-1)}_j$, and its contribution 
is related to another null hypersurface, where the cancellation we seak for works in the same way. The third term is $L$ independent because $g_{\mu\nu} \propto L^2$. The first term contribution explicitly shown has the same form as eq.\eqref{eq:renorm}, and they combine to get
\eq{
	I_{c,i} = \frac{1}{8\pi G}\log\frac{\alpha_i}{c}\left(V_{\Sigma^{(D-2)}_{i1}} - V_{\Sigma^{(D-2)}_{i2}} \right).
}
These are contributions from null hypersurfaces featured by the factor $\log(\alpha_i/c)$.

% Note that this ambiguity of scale is actually a choice of counter term, while the action with given counter term is indeed reparameterization invariant. It is similar to the situation of renormalization in quantum field theory: physics is independent of the fake parameter $\mu$ according to the Callen-Symanzik equation (reparameterization invariant), but the physical parameter should run with the scale $\mu$ (the complexity depends on $c$). 

To sum up, the total reparameterization invariant action is, 
\eq{
	I_{\rm grav} = I_{\rm bulk} + I_{\rm YGH} + I_{\rm null} + I_{\rm joint} + \Delta I,
}
where we can always set $I_{\rm null} = 0$ by choosing affine parameters on null hypersurfaces. In our case of 3d kink in pure $AdS_4$ spacetime, only null codim-1 hypersurfaces are involved as shown in Fig.~\ref{fig:calV}, thus $I_{\rm YGH} = 0$ as well. Therefore we only need to concretely compute the three contributions $I_{\rm bulk}$, $I_{\rm joint}$ and $\Delta I$.

\subsection{Bulk contributions}

Let us  first consider the bulk action, which for empty AdS space is proportional to the spacetime volume:
\eq{
	I_{\rm bulk} = \frac{1}{16\pi G} \int_{\mathcal{V}}\sqrt{-\det g}\ (R - 2\Lambda) = -\frac{3}{8\pi GL^2} \int_{\mathcal{V}}\sqrt{-\det g}.
}

It is convenient to write the metric in the new coordinate system $\xi^a = \{\zeta,\eta, \rho, \theta\}$ introduced in \eqref{eq:new_cor} and \eqref{eq:cor_trans}, 
\eq{
	g^{\xi}_{ab} = \frac{\partial\mathcal{V}^{\mu}}{\partial\xi^a}\frac{\partial\mathcal{V}^{\nu}}{\partial\xi^b}g^0_{\mu\nu}
}
where $g^0_{\mu\nu} = (L^2/z^2)\eta_{\mu\nu}$ is the original Cartesian coordinates of the Poincare patch of AdS$_4$. The volume form can thus be obtained as
\eqs{
	\sqrt{-\det g^{\xi}} = \frac{(1+h^2)H_\pm(\rho h \mp 2\zeta)}{K^2h^5(\rho h \mp \zeta - \eta)^4}L^4,
	% \sqrt{-\det g_>^{\xi}} = \frac{(1+h^2)H_-(\rho h + 2\zeta)}{K^2h^5(\rho h + \zeta - \eta)^4}L^4.
}
where upper sign is for convex kink and lower sign is for concave kink.

Using the range of parameters specified in eq(\ref{eq:range}), we get
\eq{\label{eq:bulk_preint}
	I_{\rm bulk} &\equiv -\frac{3}{8\pi GL^2}\int_{A_\delta} 2\int_{0}^{\zeta_*}\diff\zeta\int_0^{\frac{\eta_*}{\zeta_*}\zeta}\diff\eta\ \sqrt{-\det g^{\xi}}, \\
}
the factor of 2 for $\mathcal{V}_\pm$. Explicitely, the $\zeta$ and $\eta$ integrations give 
\eq{
	I_{\rm bulk} &= -\frac{L^2}{8\pi G}\int_{A_\delta}\left[Kh^5H_\pm\rho\delta^2(\rho+\rho h^2\pm Kh\delta)\right]^{-1} \\
	&\quad \times (1+h^2)^2(\rho h-\delta)^2\left[h(1+h^2)H_\mp\rho^2 
		+ 2(H_\pm^2 - 2K^2h^4)\rho\delta \pm Kh(3H_+ + 2Kh^2)\delta^2 \right],
}
where $A_\delta$ denotes the range of  $(\rho, \theta)$ as given in \eqref{eq:range} over which we integrate and the subscript $\delta$ reminds us that the range is cutoff dependent.
The $A_\delta$  integration can be written explicitly as
\eq{\label{eq:bulk_spatial_range}
	\int_{A_\delta} &= \int_{\delta/R}^{h_0}2\omega(h)\diff h\int_{\delta/h}^R\diff\rho 
}
where $\omega(h)$ is defined in \eqref{eq:def_omega}.
We perform all the $A_\delta$ integration in the Appendix~\ref{ap:int}.
% The $\rho$ integration can  be carried out analytically. Using the technique developed in the Appendix~\ref{ap:int}, we evaluate the integration in $A_\delta$ to get %the divergence structure \eqref{eq:bulk}. 
% 	\eq{\label{eq:bulk}
% 	I_{\rm bulk}^{\rm (div)} &= \int_{\delta/R}^{h_0}\ f_{\rm div}(h,\delta) - \frac{L^2}{4\pi G}\int_0^{\delta/R}\diff h\ \mathcal{B}_2(0)\frac{R^2}{\delta^2} \\
% 	& \qquad + \frac{L^2}{4\pi G}\int_0^{h_0}\diff h\ \left[ \mathcal{B}_2(h)\frac{R^2}{\delta^2} + \mathcal{B}_1(h)\frac{R}{\delta} + \mathcal{B}_0(h)\log\left[\frac{R}{\delta}\right] \right] + \mathcal{O}(\delta^0)\\
% 	&= \frac{L^2}{4\pi G}\left[ \left(\int_0^{h_0}\diff h\ \mathcal{B}_2(h)\right)\frac{R^2}{\delta^2} + \left(3 + \int_0^{h_0}\diff h\ \mathcal{B}_1(h)\right)\frac{R}{\delta} + \left(-\frac{3}{h_0} + \int_0^{h_0}\diff h\ \mathcal{B}_0(h)\right)\log\left[\frac{R}{\delta}\right] \right],
% }
% where the integrands for the numerical integrations are
% \eq{\label{eq:Bs}
% 	& \mathcal{B}_{2}(h) = -\frac{(1+h^2)H_\mp}{2Kh^2H_\pm}\omega(h), \\
% 	& \mathcal{B}_{1}(h) = \mp\frac{4(1+h^2)}{hH_\pm}\omega(h), \\
% 	& \mathcal{B}_{0}(h) = -\frac{3}{h^2}\left( 1 - \frac{1+h^2}{Kh^2}\omega(h) \right) \pm \frac{3(1+h^2)^2 - K^2h^4}{h^2(1+h^2)H_\pm}\omega(h),
% }
% and $H_\pm,\, K,$ and $\omega(h)$ are defined in \eqref{eq:Hpm},\eqref{eq:K} and  \eqref{eq:def_omega} respectively. Recall that $h(\theta)$ is the function that defines the HRT surface \eqref{eq:h_RT}.

% 
% 
% 
% 
\subsection{Boundary contributions}
The boundary of the spacetime region we are interested in consists of null codimension-1 hypersurfaces $\mathcal{W}^{\pm}$ and $\mathcal{E}^{\pm}$. 
After choosing affine parametrizations, we set the YGH term to vanish. Thus, the only term left is the $I_{\rm joint}$, which comes from three joint surfaces (see definitions in Sec.\ref{sec:setup}),
\eq{
	I_{\rm joint} = I_{W} + 2I_{J} + I_{E}
}
where $I_{J}\equiv I_{J^+}=I_{J^-}$ due to symmetry. In terms of affine parameters, the null generators of the four hypersurfaces are
\eq{
	k_{W\pm,\mu} &= \alpha g^0_{\mu\nu}U^{\nu}; \\
	k_{E\pm,\mu}^{\stackrel{<}{>}} &= -\beta g^0_{\mu\nu}(V^{\stackrel{<}{>}}_{\pm})^\nu(\lambda),
}
where $\alpha,\beta$ are positive normalization factors, and the signs are chosen so that these are outward pointing one forms. The integrands of these joint terms are respectively
\eqs{\label{eq:a_I}
	a_{W} &= -\log\left[\frac12 k_{W+,\mu} g^{0,\mu\nu} k_{W-,\nu})\right] = -\log\frac{\alpha^2\delta^2}{L^2} ,\\
	a^{\stackrel{<}{>}}_{J} &= \log\left[\frac12 k_{W\pm,\mu} g^{0,\mu\nu} k_{E\pm,\nu}^{\stackrel{<}{>}} \right] = \log\frac{\alpha\beta}{2L^2} + \bar{a}^{\stackrel{<}{>}}_{J} ,\label{eq:a_J}\\
	\bar{a}^{\stackrel{<}{>}}_{J} &= \log\frac{(\frac{\rho}{\delta}(1+h^2) \pm Kh)^2}{KH_\pm} ,\\
	a_{E} &= -\log\left[\frac12 k_{E+,\mu} g^{0,\mu\nu} k_{E-,\nu}\right] = -\log\frac{\beta^2}{L^2} + \bar{a}_{E} ,\\
	\bar{a}_{E} &=  -\log\frac{\rho^2(1+h^2)^2}{K^2h^2}.
}
where we separate the $\log L$ dependent part and $\log L$ independent part, the latter denoted by $\bar{a}$. There is no $\bar{a}_W$. This separation is inspired by the discussion in Sec.\ref{sec:ct}, which implies that the $\log L$ dependent part would be cancelled by a counter-part in the $\Delta I$ contribution. We postpone the details to the next section, and the $\log L$ independent result we find is thus
\eq{\label{eq:Jint_rhoh}
	I_{\rm joint} = \frac{1}{8\pi G}\int_{A_\delta}\sum_{X\in\{J^\pm,E\}}\bar{a}_X\sqrt{\det g_X},
}
where $g_X$ is given in Appendix~\ref{ap:geo} and the details of the integrations are presented in Appendix~\ref{ap:int}. 

\subsection{Counter term contributions}
As discussed in Sec.\ref{sec:ct}, we have to include the counter term \eqref{eq:ct} to restore the null surface reparameterization invariance. The ingredients of the integrations are all provided in Appendix~\ref{ap:geo}. 
As shown in Sec.\ref{sec:ct}, $\Delta I$ can be separated into two parts, one proportional to $\log L$, and the other independent of $L$. The $\log L$ contributions combine with the joint terms to get
\eq{\label{eq:hyper}
	I_c &= 2I_{c,\alpha} + 2I_{c,\beta} \\
	&= \frac{1}{4\pi G}\left[ V_W\log\frac{c}{\alpha} - V_J\log\frac{2c^2}{\alpha\beta} + V_E\log\frac{c}{\beta} \right]
}
where the $\log 2$ in the second term stems from eq.\eqref{eq:a_J}. We use the Ryu-Takayanagi formula
\eq{\label{eq:RT}
	S_{\rm EE}(\mathcal{A}) &= \frac{1}{4G}V_E = \frac{L^2}{4G}\left[\frac{2R}{\delta} - s(h_0)\log\frac{R}{\delta}\right] ,\\
	& s(h_0) = \frac{2}{h_0} + \int_0^{h_0}\frac{2\diff h}{h^2}\left(1-\frac{1+h^2}{Kh^2}\omega(h)\right).
}
and rearrange the terms to get
\eq{\label{eq:Ic}
	I_c = \frac{L^2}{4\pi G}\frac{V_{\delta}}{\delta^2}\log\frac{c}{\alpha} - \frac{S_{\rm EE}}{\pi}\log\frac{2c}{\alpha} + \frac{V_E - V_J}{4\pi G}\frac{2c^2}{\alpha\beta}.
}
where we also define
\eq{
	V_{\delta} &= \frac{V_W}{\sqrt{\det g_W}} = \frac{\delta^2}{L^2}V_W \\
	&= V_{\mathcal{A}} + \mathcal{O}(\delta)
}
whose leading contribution in $\delta$ equals the volume of the boundary region $\mathcal{A}$\footnote{As the volume of $\mathcal{A}$ heavily depends on the choice of IR cutoff, the exact form of it is not important, so we do not present it here.}. Hence the first term in eq.\eqref{eq:Ic} is a leading volume law contribution. The second term, as written explicitly, is proportional to the entanglement entropy of $\mathcal{A}$, and is an area law contribution. 
The third term in \eqref{eq:Ic} can be proved to be subleading, as the leading area law terms in $V_J$ and $V_E$ cancel each other. The structure of $I_c$ derived above is general.

For the $L$ independent part, we use the formula \eqref{eq:ct} to obtain
\eq{\label{eq:ct_calc}
	& \Delta I = 2\Delta I_{\mathcal{E}} + 2\Delta I_{\mathcal{W}} ,\\
	& \Delta I_{\mathcal{E}} = -\frac{1}{8\pi G}\int_{A_\delta}\int_0^{\lambda_*}\diff\lambda\sqrt{\det g(\lambda)}\ \Theta(\lambda)\log\frac{\delta|\Theta(\lambda)|}{2} ,\\
	& \Delta I_{\mathcal{W}} = -\frac{1}{8\pi G} \int_{A_\delta}\int_0^{\tilde\lambda_*}\diff\lambda\sqrt{\det \tilde{g}(\tilde\lambda)}\ \Theta(\tilde\lambda)\log\frac{\delta|\Theta(\tilde\lambda)|}{2},
}
with the expansion rate $\Theta$ given by \eqref{eq:Theta_E} and \eqref{eq:Theta_W},
while $\lambda$ and $\tilde\lambda$ are null affine parameters on $\mathcal{E}$ and $\mathcal{W}$, $\lambda_*$ and $\tilde\lambda_*$ are their end values given by \eqref{eq:lambdaE_star} and \eqref{eq:lambdaW_star}. The integrations are performed in Appendix~\ref{ap:int}.

% 
% Now let's turn to the $L$ independent part of $\Delta I$. 
% %
% We perform the $\lambda$ integration first:
% \eq{\label{eq:ct_beta}
% 	& \int_0^{\lambda_*}\diff\lambda\sqrt{\det g(\lambda)}\ \Theta(\lambda)\log\frac{\delta|\Theta(\lambda)|}{2} \\
% 	= & -\frac{\delta^2(1+h^2)}{2K\rho h^4}\Big[ \frac{K^2h^2(\rho h - \delta)^2}{(\rho(1+h^2) + Kh\delta)^2} \left(1+2\log\frac{2Kh^2(\rho h-\delta)(\rho(1+h^2)+Kh\delta)}{\delta H_+(\rho H_- + 2Kh\delta)}\right) \\
% 	& \qquad \qquad + 2\log\left(1-\frac{K^2h^2(\rho h-\delta)^2}{(\rho(1+h^2)+Kh\delta)^2}\right) \Big]
% }
% for $\mathcal{E}$, and
% \eq{\label{eq:ct_alpha}
% 	& \int_0^{\tilde{\lambda}_*}\diff{\tilde\lambda}\sqrt{\det \tilde{g}(\tilde\lambda)}\ \Theta(\tilde\lambda)\log\frac{\delta|\Theta(\tilde\lambda)|}{2} \\
% 	= & -\frac{\delta^2(1+h^2)(\rho H_- + 2Kh\delta)}{2K h^4} \\
% 	& \times \left[ \frac{h^2}{H_+\delta^2}(1+\log\frac{2\delta}{L}) - \frac{H_+}{(\rho(1+h^2) + Kh\delta)^2}\left(1 + 2\log\frac{2h(\rho(1+h^2)+Kh\delta)}{\delta H_+}\right) \right]
% }
% for $\mathcal{W}$. Next, we use the procedure described in section~\ref{ap:int} to do the integrations eq.\eqref{eq:bulk_spatial_range}, and extract the divergence structures.

%%%%%%%%%%%%%%%%%%%%%%%%%%%%%%%%%%%%%%%%%%%%%%%%%%%%%%%%%%%%%%%%

\section{Final result and discussion}
\label{sec:result}

%With the relations given in eq(\ref{eq:relation}), we have 6 independent numerical integrations with integrand $\tilde{\mathcal{J}}^W_{2}$, $\tilde{\mathcal{J}}^W_{1}$, $\mathcal{J}_{0}$, $\mathcal{B}_{0}$, $d_{\mathcal{A}0}$, $d_{\mathcal{B}0}$. The behaviors of these integrations will be shown in the next sections. 
Adding the contributions from bulk and boundary, the total action is obtained in \eqref{eq:I_total},
\eq{\label{eq:I_total_main}
	I_{\rm grav} & = I_c + \frac{L^2}{4\pi G}\left[\frac{\Delta(h_0)R}{\delta} + \ell(h_0)\log\frac{R}{\delta}\right] \\
	& = \frac{L^2}{4\pi G}\frac{V_{\delta}}{\delta^2}\log\frac{c}{\alpha} - \frac{S_{\rm EE}}{\pi}\log\frac{2c}{\alpha} \\
	& + \frac{L^2}{4\pi G}\left[ \Delta(h_0)\frac{R}{\delta} + \left(\ell(h_0) + \ell'(h_0)\log\frac{2c^2}{\alpha\beta} \right)\log\frac{R}{\delta} \right] + \mathcal{O}(\delta^0),
}
$V_\delta$ is the volume of $\mathcal{A}$ up to higher order corrections, and $S_{\rm EE}$ is the entanglement entropy of $\mathcal{A}$ given in \eqref{eq:RT}. The details of the coefficient functions $\Delta(h_0)$, $\ell(h_0)$ and $\ell'(h_0)$ are defined in \eqref{eq:def_Delta}, \eqref{eq:ell} and \eqref{eq:ellp}. 

To identify the kink contributions from these terms, we can perform a quick check by taking the flat angle limit $h_0 \to \infty$. The kink contribution should vanish at this limit. As will be shown in the rest of the section, the three coefficient functions have the following behavior
\eq{
	\lim_{h_0\to\infty} \Delta(h_0) = 1, \quad \lim_{h_0\to\infty} \ell(h_0) = 0, \quad \lim_{h_0\to\infty} \ell'(h_0) = 0.
}
The first one says that the extra area term $\Delta(h_0)$ does not come from the kink, which we will be discussing about in Sec.\ref{sec:geo}. The latter two confirm that the $\log\delta$ divergences are kink contributions, which will be analyzed in Sec.\ref{sec:univ}.

\subsection{General structure of CA subregion complexity}
\label{sec:geo}

In light of the discussions in \citep{Carmi:2016wjl}, the divergence structure of the holographic complexity of subregion $\mathcal{A}$ can be expressed in terms of volume integration in $\mathcal{A}$ and surface integration on $\partial\mathcal{A}$
\eq{\label{eq:C_general}
	\mathcal{C}_{\mathcal{A}} = \int_{\mathcal{A}} v(\mathcal{R},K) + \int_{\partial \mathcal{A}} b(\mathcal{R},\tilde{K};\mathbf{s},\mathbf{t}) + {\rm finite\ terms},
}
In the above equation, we have $\mathcal{R}$ denoting the spacetime curvature, $K$ denoting the exterior curvature of the time slice, $\tilde{K}$ denoting the exterior curvature of $\partial\mathcal{A}$, and $\mathbf{s}$, $\mathbf{t}$ denoting the spacelike and timelike normal vectors of the $\partial\mathcal{A}$. 

This expression is only for subregions with smooth surface, similar to gravitational action only with YGH term. In general, there could be higher codimension defects on the surface, like the cube example we mentioned in Sec.~\ref{sec:highCoD}, and these defects could contribute independently to the complexity. Thus the most general form would be an expression similar to \eqref{eq:I_general}. In this paper, we only deal with a kink shape in two spatial dimensions, where the only singularity on the surface is the point-like kink tip. Thus the only extra term we expect is a local contribution that does not involve any integrations. For higher dimensional subregions with singular surface, integration might be needed for contributions from non-point-like creases. 

The main point of this notation is that these integrands are local functions in $\mathcal{A}$ or on $\partial\mathcal{A}$. In particular, we can expand the integrands in powers of the UV cutoff $\delta$
\eq{
	v = \delta^{1-d}\sum_{i=0}^{d-1}v_i\delta^i, \qquad b = \delta^{2-d}\sum_{i=0}^{d-2}b_i\delta^i,
}
while $v_i$, $b_i$ has mass dimension $i$ coming from the curvatures. For highest order $i$ in the series, $\delta^0$ should be replaced by a universal contribution $\log\delta$.
In the kink case, all curvatures involved in the volume and area integrations vanish: $\mathcal{R}=0$ because we are in Poincare patch, $K = 0$ as we are on a trivial flat time slice, and $\tilde{K} = 0$ for the straight sides of the kink. Hence, we expect the integrands to be constants, and the integrations simply give the volume and surface area of $\mathcal{A}$. All the other contributions that are not proportional to the volume or area should be attributed to the kink singularity. 

\begin{figure}[!hbt]
	\centering
	\includegraphics[width=0.5\textwidth]{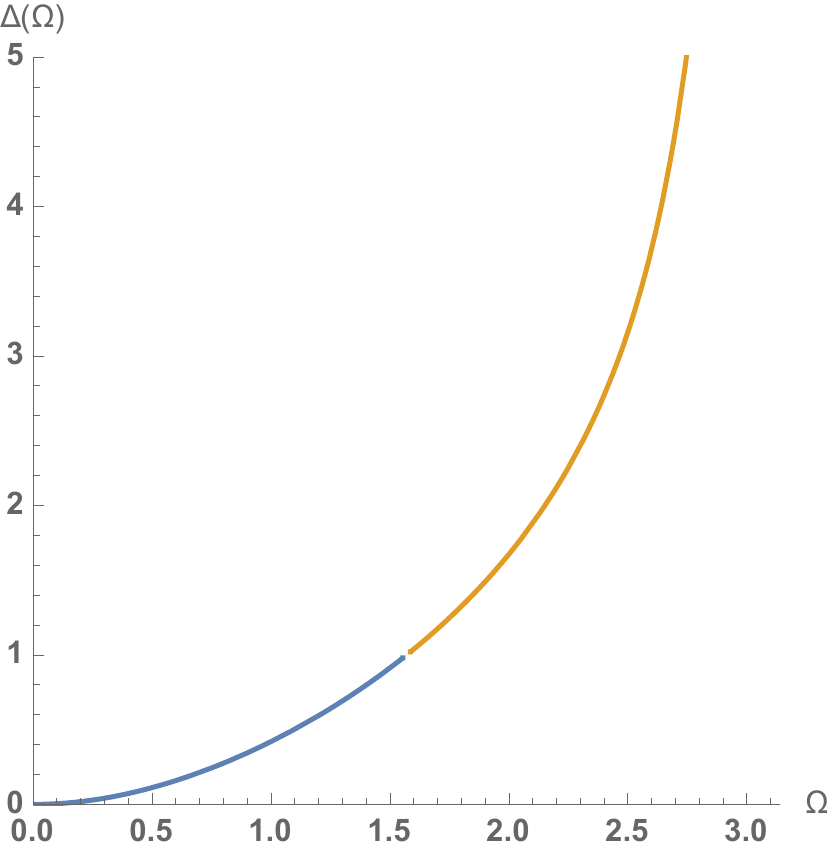}
	\caption{The $\Delta(h_0)$ function both for convex and concave kinks are presented. The blue line with $\Omega < \pi/2$ is for convex kink, while the concave kink is represented by the yellow line with $\Omega > \pi/2$. The whole curve is smooth. At flat angle we have $\Delta(\pi/2)=1$. }\label{fig:Delta}
\end{figure}

Our final result eq.\eqref{eq:I_total_main} gives
\eq{\label{eq:general_feature}
	v_0 = \frac{L^2}{4\pi G}\log\frac{c}{\alpha}, \qquad b_0 = \frac{L^2}{4\pi G}\left[ -\log\frac{2c}{\alpha} + \frac{1}{2}\Delta(h_0) \right].
}
We see that the $\Delta$ term is an exception of the structure \eqref{eq:C_general}. It is an area law term, and it depends on the opening angle through the dependence on $h_0$. From Fig.~\ref{fig:Delta}, we see that\footnote{
In Fig.~\ref{fig:Delta} we adopt a different IR cutoff that has constant radius on the boundary instead of on the RT surface. The reason is that under this choice the function grows monotonically, and we can see a clean relation with the openning angle. The choice we make in other part of the paper would induce a decrease of the function at small $\Omega$, which we think is distracting.
} $\Delta$ starts from 0 when the kink is sharp, and approaches 1 when the kink becomes smooth. For concave kinks, $\Delta$ continues growing with the openning angle. Our guess is that this contribution is related to the fact that we are dealing with a non-compact subregion: when we treat it as an approximate part of a large compact subregion, this term characterize the transition part between the kink portion and the rest. Thus we conjecture that for a normal compact subregion, eq.\eqref{eq:general_feature} without the $\Delta(h_0)$ term is a general form of leading divergent contributions to holographic complexity. 

Note that both the volume law and area law terms in eq.\eqref{eq:general_feature} depend on the ``renormalized'' parameter $\tilde{\alpha} \equiv \alpha/c$. In order for the leading volume term to be positive, we have to set $\tilde{\alpha} < 1$, which implies that the area term has to be negative. 
The positive volume law is easy to understand as complexity of state naturally grow with the size of the subregion. 
If we interpret the area law term as the entanglement with the complementary of the subregion, the negative sign can be understood as \emph{loss} of detailed information of the entanglement when the outside is traced out. If we use the definition of subregion complexity as the minimal complexity to construct a purification of the density matrix $\rho_{\mathcal{A}}$ \cite{Agon:2018zso}, the amount of complexity that can be saved during the minimization among purifications $\{ \psi | {\rm Tr}_{\bar{\mathcal{A}}}\ket{\psi}\bra{\psi} = \rho_{\mathcal{A}} \}$ should behave as the amount of entanglements, \ie~ the entanglement entropy, among which we have the freedom to setup the details. 
To get a clearer picture, let us look at the following quantity for holographic complexity
\eq{\label{eq:cplx_e}
	\mathcal{I}_{\mathcal{C}}(\mathcal{A},\mathcal{B}) \equiv \mathcal{C}(\mathcal{A}\cup\mathcal{B}) - \mathcal{C}(\mathcal{A}) - \mathcal{C}(\mathcal{B}),
}
similarly defined as mutual information $\mathcal{I}(\mathcal{A},\mathcal{B})$ for entanglement entropy. The leading behavior in \eqref{eq:Ic} then indicates that
\eq{\label{eq:cplx-mutual_info}
	\mathcal{I}_{\mathcal{C}}(\mathcal{A},\mathcal{B}) \sim \frac{1}{\pi}\log\frac{2c}{\alpha}\mathcal{I}(\mathcal{A},\mathcal{B}),
}
because the volume term exactly cancels. This proportionality can be understood as follows: On one hand, after building quantum states in subregions $\mathcal{A}$ and $\mathcal{B}$, the extra complexity needed to reproduce the state in the combined region $\mathcal{A}\cup\mathcal{B}$ is what $\mathcal{I}_{\mathcal{C}}$ means. On the other hand, the extra effort to make in this process is to build the correct entanglement between $\mathcal{A}$ and $\mathcal{B}$, the amount of which is the mutual information $\mathcal{I}(\mathcal{A},\mathcal{B})$. Therefore we call $\mathcal{I}_{\mathcal{C}}$ the \emph{complexity of entanglement}.

% We can also apply this logic to the thermal state, where the idea of holographic complexity was first motivated \cite{}. The complexity growth that is dual to the wormhole growth is completely hidden from the one-sided thermal state $\rho_{\mathcal{A},\mathcal{B}}$, but it appears in the complexity of entanglement \eqref{eq:cplx_e}, which precisely captures the growing part of the complexity. However, the relation \eqref{eq:cplx-mutual_info} does not hold anymore, because the behavior \eqref{eq:Ic} is not for cases when two sides of the black hole are both involved, like in $\mathcal{C}(\mathcal{A}\cup\mathcal{B})$.

We can take another simple case to check the general behavior we got. In \citep{Carmi:2016wjl}, general formula for spherical subregion complexity was presented for all bulk dimension $d$.
% \eq{
% 	\mathcal{C}^{\rm old}_A({\rm }) &= \frac{L^2}{4\pi^2 G}\Bigg[ \left(-\frac{1}{2}\frac{\pi R^2}{\delta^2} + \frac{3-2\log2}{2}\frac{2\pi R}{\delta} - 2\pi\log\frac{R}{\delta} \right) \\
% 	& + \log\frac{L}{\alpha\delta}\left(\frac{\pi R^2}{\delta} - \frac{2\pi R}{\delta} + \pi \right) \Bigg],
% }
However, this did not take into account the counter term contributions that we have included in the  kink calculations presented here. With the addition of the counter terms, we calculate the  complexity-action of a disk region in CFT$_3$ and obtain,
\eq{
	\mathcal{C}_A({\rm disk}) &= \frac{L^2}{4\pi^2 G}\Bigg[ \frac{\pi R^2}{\delta^2}\log\frac{c}{\alpha} - \frac{2\pi R}{\delta}\log\frac{2c}{\alpha} - 2\pi\log\frac{R}{\delta} \Bigg].
}
It reproduces the pattern  in the first line of \eqref{eq:I_total_main}: The first term is the $V_\delta$ volume  term, and the second term is a negative area  term. The $\log\delta$ term here is of course not from singular surface, but from area integration $\int_{\partial\mathcal{A}}b$, as here we have subleading term in $b_1 \sim \tilde{K}$ due to the non-vanishing curvature $\tilde{K} = 1/R$. % Whether universal formula for such subleading terms exists or not is an interesting question.

\subsection{Universal contributions from the the kink}
\label{sec:univ}

In this section, we focus on the universal $\log\delta$ divergences in \eqref{eq:I_total_main}. The entropy term contains a universal contribution $s(h_0)$, which is inherited from the entanglement entropy $S_{\rm EE}({\rm kink})$ given in \eqref{eq:RT}. The new universal terms are the $\ell(h_0)$ and $\ell'(h_0)$, both coming from the kink feature. Thus we have the total universal contributions as
\eq{
	I^{\rm univ} = \frac{L^2}{4\pi G}\left[ s(h_0)\log\frac{2c}{\alpha} + \ell'(h_0)\log\frac{2c^2}{\alpha\beta} + \ell(h_0)\right]\log\frac{R}{\delta}.
}

We first look at the simpler one $\ell'(h_0)$ defined in \eqref{eq:ellp}. This contribution does not show up in the spherical subregion computation in \citep{Carmi:2016wjl}, because the RT surface for spherical region is a bifurcating Killing horizon, which implies the vanishing expansion rate $\Theta=0$ on the hypersurface $\mathcal{E}$. In other words, volumes of spatial slices are the same on $\mathcal{E}$, and hence $V_E = V_J$ and $\ell' \sim V_E - V_J = 0$. 
%
% \todo{
% This relation inspires us to link the origin of this contribution to the locality of the modular Hamiltonian, as it is natural to expect higher complexity due to non-locality \cite{Couch:2017yil}. As is well known, modular Hamiltonian is local \cite{Casini:2011kv} for spherical subregion in CFT$_d$:
% \eq{
% 	H_{\mathcal{A}} = 2\pi\int_{\mathcal{A}}\Diff{x}{d-1}\frac{R^2-r^2}{2R}T^{00}(x) + c'
% }
% The complexity contribution $\ell'$ characterizes the deviation from this behavior. Using focusing theorem, it is easy to show that $\ell'=0$ is a necessary and sufficient condition of a vanishing expansion rate $\Theta = 0$ on $\mathcal{E}$ at least between $E$ and $J$, while non-zero $\Theta$ developed after the joint $J$ can be attributed to external disturbance on the boundary after the time of the state we analyse the complexity of, due to the definition of the WdW patch. In sum, we conjecture that $\ell'$ could characterize the non-locality of the modular Hamiltonian of a subregion at a given time (at which we compute the complexity and the $\ell'$). Hence we name it the \emph{complexity of non-locality}.
% }
% 
% \newpage
% 
% \hrule
% \todo{
Using focusing theorem $\dot{\Theta}\le 0$, it is easy to show that $\ell'=0$ is a necessary and sufficient condition for a vanishing expansion rate, $\Theta = 0$, on $\mathcal{E}$\footnote{Non-zero values of $\Theta$ on $\mathcal{E}$ may be developed after the joint $J$, but since $J$ is at the boundary of the WdW patch, these non-vanishing $\Theta$ are not relevant for the quantum state that we compute holographic complexity for.}. On the other hand, a spherical subregion in CFT$_d$ is one of the few cases where  the  modular Hamiltonian is local \cite{Casini:2011kv} and, as we mentioned above, in this case $\Theta=0$. Thus, it is tempting to conjecture that the  boundary meaning  of $\ell'$  is related to  the non-locality of the modular Hamiltonian of a subregion at a given time. Note that a higher complexity due to non-locality is to be expected according to \cite{Couch:2017yil}. Understanding if $\ell'$ indeed represents the \emph{complexity of non-locality} is a question that deserves further study and that we leave for future work.
% }
% \hrule

\begin{figure}[!hbt]
	\centering
	\includegraphics[width=0.4\textwidth]{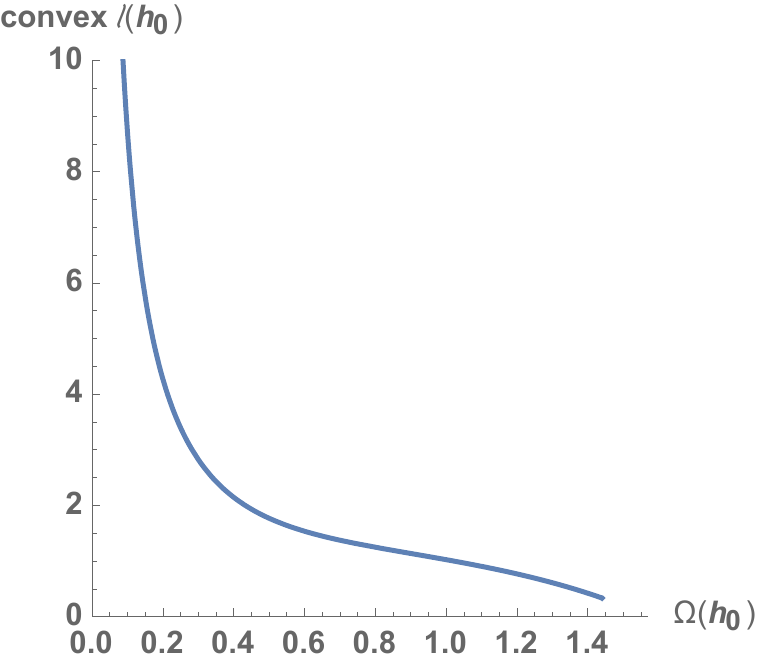}
	\includegraphics[width=0.4\textwidth]{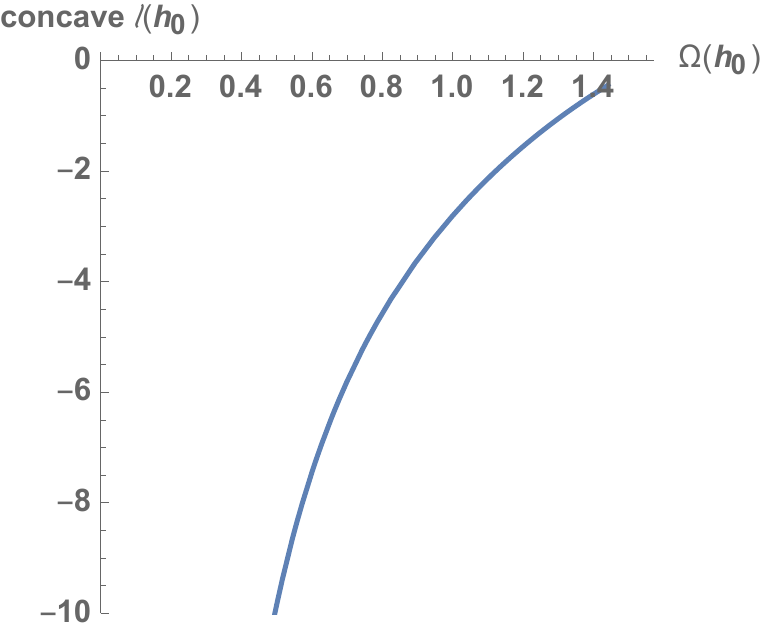}
	\caption{The $\ell(h_0)$ function both for convex (left panel) and concave (right panel) kinks are presented separately. Both of them vanish at the flat angle limit ($\Omega = \pi/2$). }\label{fig:ell}
\end{figure}

The function $\ell(h_0)$ is plotted in Fig~\ref{fig:ell}. The two panels present convex and concave kinks. 
From the expression \eqref{eq:ell}, we find that except for the last term, all the other terms are symmetric for convex and concave cases, which means it is an even function with respect to the flat angle $\Omega = \pi/2$. The last term, coming from the bulk contribution, is an almost odd function, and dominates over the even terms. Thus near the flat angle, the bulk contribution has the dominate behavior
\eq{
	\lim_{h_0\to\infty}\ell(h_0) = \frac{3\pi}{2h_0} = 3\left(\frac{\pi}{2} - \Omega\right).
}
The last equation comes from \eqref{eq:open_angle}. This linear behavior around the flat angle also dominates quadratic behaviors for $s(h_0)$ and $\ell'(h_0)$. Therefore, similar to the CV result \eqref{eq:CV_result}, the subregion-CA kink contribution also linearly depends on the deviation from the flat angle limit.

%--------------------- Sec 6 ------------------------------------------------------------------------
\section{Conclusions and future directions}
\label{sec:conclusions}

%Inspired by the holographic entanglement entropy studies for boundary subregions with singular surface \citep{Bueno:2015rda,Bueno:2015xda,Bueno:2015lza}, and similar computations of complexity-volume for the same subregions \citep{Bakhshaei:2017qud}, 
We studied complexity-action for a particular singular subregion, {\it i.e.} the kink in CFT$_3$. The calculation is quite involved and great care was taken to develop appropriate techniques that can be generalized to other singular and higher dimensional regions. The result we obtain for subregion-CA  consists of a volume term, a negative area term and cutoff independent divergent terms of order $\log\delta$ coming from the kink singularity. 

The concise form of the final result presented in \eqref{eq:I_total_main}  is in great part due to the effect of the boundary counter term that is introduced to restore reparametrization invariance of the null boundaries. This counter term produces several non-trivial cancellations, simplifies greatly the divergence structure and leads to a final result with no $\log^2\delta$ terms. We identified a general form of null-norm dependent contributions \eqref{eq:Ic}, and also studied the limiting behavior of the cutoff independent contribution.

Let us make a couple of observations regarding the possible contribution from entanglement to complexity-action. 
We find a negative area contribution to subregion complexity. If we relate this area term with entanglement this negative contribution can intuitively be understood as the information lost when we trace out the complement. The less information contained in the state, the easier (with less complexity) can we construct the state from quantum circuit.
% However, we cannot verify this quantitatively, because the contribution from entanglement might  not be proportional to the entanglement entropy\footnote{As an example suppose there is a Bell pair separated by $\partial A$. The entanglement entropy would always count it as $\log2$, while the complexity saved by tracing out $\bar{A}$ depends on the distance between the pair, due to the locality of gates. The dependence can in general be complicated in terms of the Hamiltonian and choice of gates.}. 
% In general entanglement might be  weighted differently in entanglement entropy than in  complexity. 
Perhaps more importantly, the area term contribution we find  points to a robust difference between subregion-CA and subregion-CV. An area law term in CV is prohibited in a time-symmetric configuration \cite{Carmi:2016wjl}. However, the contribution discussed above should exist universally, including the time-symmetric case. {\it Thus,  the area term that we find can only be obtained in CA not CV.} 

To conclude the discussion, we list some future directions related to our work:  
\begin{itemize}
		\item In the case of a kink singularity studied here , the  caustics are always outside region $\mathcal{V}$ and thus, played no role in our calculation (Section  \ref{sec:caustic}). However, for more general singular regions this need not be the case, caustics might occur  on the relevant part of $\mathcal{E}$. It is not known  if the caustic itself provides additional contributions to the complexity. Due to the vanishing of the induced metric at the caustic, it would be either a higher co-dimensional singular feature, or a null joint. Action contributions for both cases are unknown and worth studying. The physical meaning of such cases is also interesting to explore.
		\item Methods of computing action contributions from higher co-dimensional surfaces are also necessary for, say, polyhedral corners on $\partial\mathcal{V}$. 
	%	\item We have seen that the IR cutoff we chose brought into an area law term that depends on the openning angle through the function $\Delta(h_0)$. This contribution cannot be interpretted with geometrical origin, and thus is not theoretically preferable. We adopted this IR cutoff due to computational convenience, but new methods could be found for convenient computation of other IR cutoff. We hope that for some better IR cutoff, the angle dependent area law term can be eliminated, and some universal area law coefficients can be verified.
%	\end{enumerate}
	
	\item The connection found between complexity contribution and entanglement is worth exploring. More detailed studies on such contributions in holographic complexity and the implication from certain field theory definition of complexity are interesting directions to pursue. %The fact that such features do not appear in CV also provides a criterion for the validity of the two holographic approaches.
	
%	\item Example of finite subregion with kink should be studied to verify the locality of the kink contributions, {\it i.e.} there should be \emph{simple sum} of characteristic contributions from all kinks or other singular features on the surface in the complexity/action. 
	
	\item We can modify the boundary state or the theory to see how the cutoff independent terms change. In the study of entanglement entropy, certain universality of the coefficient functions was found, namely the ``corner charges''. It would be nice if subregion complexity also has some universality that one can use to reveal information of the underlying CFT.
	
	\item Similar computation can be extended to higher dimensional singular surfaces, including smooth cones, polyhedral cones and their product with flat dimensions (creases). It will be interesting to see the dependence of the ``corner charges'' on dimensionality. New divergence structures are also expected.
\end{itemize}

%%%%%%%%%%%%%%%%%%%%%%%%%%%%%%%%%%%%%%%%%%%%%%%%%%%%%%%%%%%%%%%%%%

\acknowledgments
We would like to thank Cesar Ag\'on and  Matthew Headrick for useful comments. We also  thank the anonymous referee for her/his valuable comments. 
This material is based upon work supported by the National Science Foundation under
grants PHY-1620610 and PHY-1820712. E.C. also thanks the  Aspen Center for Physics, supported by National Science Foundation grant PHY-1607611, where part of this work was performed.

%%%%%%%%%%%%%%%%%%%%%%%%%%%%%%%%%%%%%%%%%%%%%%%%%%%%%%%%%%%%%%%%%%%%%%
\appendix
	%--------------------- App A ------------------------------------------------------------------------
	
	\section{Details of the action computation}
	\label{ap:int}
	%We adopt the following procedure to do $\theta$ integration over the range $(\delta/R, h_0)$ and extract the UV divergent pieces. 

	We adopt the following procedure to do $\rho$ and $\theta$ integrations of a general integrand $f_0$ over the range defined by $A_{\delta}$. Using eq(\ref{eq:bulk_spatial_range}), we obtain
	\eq{
		F = \int_{A_{\delta}}\ f_0(\rho,h(\theta),\delta)
		= \int_{\delta/R}^{h_0}2\omega(h)\diff h \int_{\delta/h}^R \diff \rho\ f_0(\rho,h,\delta).
	}
	The $\rho$ integration is usually carried out analytically, so we can define
	\eq{
		f_1(h,\delta) = 2\omega(h)\int_{\delta/h}^R \diff \rho\ f_0(\rho,h,\delta).
	}
	Then we series expand $f_1(h,\delta)$ in powers of $h$, and extract the part that is divergent at $h=0$
	\eq{
		f_1(h,\delta) = f_{\rm div}(h,\delta) + f_2(h,\delta).
	}
	As $f_{\rm div}$ is in power series of $h$, its integration can be carried out explicitly
	\eq{
		F_h(\delta) = \int_{\delta/R}^{h_0}\ f_{\rm div}(h,\delta) = \sum_{i>0}F_h^{(i)}\delta^{-i} + F_h^{(0)}\log\delta + \mathcal{O}(\delta^0),
	}
	while we do integration of $f_2(h,\delta)$ for each divergent powers of $\delta$ numerically. Namely, for the expansion
	\eq{
		f_2(h,\delta) = \sum_{i>0}f_2^{(i)}\delta^{-i} + f_2^{(0)}(h)\log\delta + \mathcal{O}(\delta^0),
	}
	we get
	\eq{\label{eq:F^i}
		F^{(i)} &= F_h^{(i)} - \sum_{j>0}\left[ \int_0^{\delta/R} \diff h\ f_2^{(i+j)}(h) \right]\Bigg|_{\mathcal{O}(\delta^{j})} + \int_0^{h_0}\diff h\ f_2^{(i)}(h)\\
		&= \left( F_h^{(i)} - \sum_{j>0}R^{-j}\frac{{\rm d}^{j-1}f_2^{(i+j)}}{{\rm d}h^{j-1}}\Big|_{h=0} \right) + \int_0^{h_0}\diff h\ f_2^{(i)}(h)
	}
	for $i$th order divergence. Note that although the regularity at $h=0$ guarantees that there's no new divergence from integration at $h=0$, it can still produce weakend divergence from the second term above. For instantce
	$\int_0^{\delta/R}\diff h\ f_2^{(2)}(h)$
	may contribute to $F^{(1)}$ if $f_2^{(2)}(0)\neq 0$. In sum, the UV divergent part of the integral $F$ is
	\eq{
		F_{\rm div} = \sum_i F^{(i)}\delta^{-i}
	}
	
	There is a subtlety in this procedure. Although $f_2(h,\delta)$ is regular at $h=0$ by design, after expansion in $\delta$, it may not be regular order by order. It depends on the interchangability of the $h\to0$ limit and $\delta\to0$ limit. In the present case the coefficients $f_2^{(i)}(h)$ are all integrable at $h=0$. 
	%\todo{study when this technique breaks down and whether there is physical meaning here.}

	Now we apply this procedure to the real calculations. In what follows, the functions $ \omega (h),\, H_\pm (h) $  and the integration constant $K$ appear frequently in our calculations. They are defined in equations \eqref{eq:def_omega} ,  \eqref{eq:Hpm} and \eqref{eq:K}  respectively. For sake of completness we list them again here:
	\begin{align}
&	K = \frac{1+h^2}{h^2\sqrt{1+h^2+h^{\prime 2}}} = \frac{\sqrt{1+h_0^2}}{h_0^2}\\
& \omega(h)=\frac{Kh^2}{\sqrt{(1+h^2)(1+h^2-K^2h^4)}}\\
&	H_\pm(h)\equiv 1+h^2\pm Kh^2
\end{align}
	
%%%%%%%%%%%%%%%%%%%%%%%%%%%%%%%%%%%%%%%%%%%%%%%%%%%%%%%%%%%%%%%%%
\subsection{Bulk contributions}
	First look at the bulk contribution eq(\ref{eq:bulk_preint}):
	\eq{
		f_0(\rho,h,\delta) &= -\frac{L^2}{8\pi G} \frac{(1+h^2)^2(\rho h-\delta)^2}{Kh^5H_\pm\rho\delta^2(\rho+\rho h^2\pm Kh\delta)} \\
		& \quad \times \left[h(1+h^2)H_\mp\rho^2 + 2(H_\pm^2 - 2K^2h^4)\rho\delta + Kh(3H_\pm + 2Kh^2)\delta^2 \right].
	}
	Hereafter, the subscripts $\pm$ are for convex and concave kinks respectively.
	The $\rho$ integration can be done analytically to get $f_1(h,\delta)$, which decomposes as
	\eq{
		&f_1(h,\delta) = f_{\rm div}(h,\delta) + f_2(h,\delta),\\
		&f_{\rm div}(h,\delta) = \frac{L^2}{4\pi G}\left[ \frac{2\delta}{Rh^3} + \frac{1}{h^2}\left(3\log\frac{Rh}{\delta}-\frac{3}{2}-\frac{K\delta^2}{2R^2}\right) + \frac{2K\delta}{Rh} \right], \\
		&f_2(h,\delta) = \frac{L^2}{4\pi G}\left[ \mathcal{B}_2(h)\frac{R^2}{\delta^2} + \mathcal{B}_1(h)\frac{R}{\delta} + \mathcal{B}_0(h)\log\left[\frac{R}{\delta}\right] + \mathcal{O}(\delta^0) \right].
	}
	Using the notation developed above we get the divergence coefficients as
	\eq{
		& F^{(2)}_{\rm bulk} = \frac{L^2R^2}{4\pi G}\int_0^{h_0}\diff h\ \mathcal{B}_2(h) ,\\
		& F^{(1)}_{\rm bulk} = \frac{L^2R}{4\pi G}\left[3 + \int_0^{h_0}\diff h\ \mathcal{B}_1(h) \right] ,\\
		& F^{(0)}_{\rm bulk} = -\frac{L^2}{4\pi G}\left[-\frac{3}{h_0} + \int_0^{h_0}\diff h\ \mathcal{B}_0(h) \right].
	}
%	Note that the non-zero limit of $\mathcal{B}_{2}$ at $h=0$ will contribute to the $\delta^{-1}$ divergence after integrated against the UV cutoff region, as indicated by the second term in eq(\ref{eq:F^i}). Overall, the UV divergences are:
%	\eq{\label{eq:bulk}
%		I_{\rm bulk}^{\rm (div)} &= \int_{\delta/R}^{h_0}\ f_{\rm div}(h,\delta) - \frac{L^2}{4\pi G}\int_0^{\delta/R}\diff h\ \mathcal{B}_2(0)\frac{R^2}{\delta^2} \\
%		& \qquad + \frac{L^2}{4\pi G}\int_0^{h_0}\diff h\ \left[ \mathcal{B}_2(h)\frac{R^2}{\delta^2} + \mathcal{B}_1(h)\frac{R}{\delta} + \mathcal{B}_0(h)\log\left[\frac{R}{\delta}\right] \right] + \mathcal{O}(\delta^0)\\
%		&= \frac{L^2}{4\pi G}\left[ \left(\int_0^{h_0}\diff h\ \mathcal{B}_2(h)\right)\frac{R^2}{\delta^2} + \left(3 + \int_0^{h_0}\diff h\ \mathcal{B}_1(h)\right)\frac{R}{\delta} + \left(-\frac{3}{h_0} + \int_0^{h_0}\diff h\ \mathcal{B}_0(h)\right)\log\left[\frac{R}{\delta}\right] \right].
%	}
	where the numerical integrands are given by
	\eq{\label{eq:Bs}
		& \mathcal{B}_{2}(h) = -\frac{(1+h^2)H_\mp}{2Kh^2H_\pm}\omega(h), \\
		& \mathcal{B}_{1}(h) = \mp\frac{4(1+h^2)}{hH_\pm}\omega(h), \\
		& \mathcal{B}_{0}(h) = -\frac{3}{h^2}\left( 1 - \frac{1+h^2}{Kh^2}\omega(h) \right) \pm \frac{3(1+h^2)^2 - K^2h^4}{h^2(1+h^2)H_\pm}\omega(h).
	}
	
%%%%%%%%%%%%%%%%%%%%%%%%%%%%%%%%%%%%%%%%%%%%%%%%%%%%%%%%%%%%%%%%%
\subsection{Joint contributions}
\label{sec:joint}

	Next, we look at the joint contributions. For the $L$ independent part given by eq(\ref{eq:Jint_rhoh}), we have the $\rho$ integration results:
	\eq{
	f_1(h,\delta) &\equiv \frac{2\omega(h)}{8\pi G}\int_{\delta/h}^{R}\diff\rho\ 2\bar{a}_{J}\sqrt{\det g_{J^\pm}} + a_E\sqrt{\det g_E} \\
	&= \frac{L^2}{4\pi G}\left[ \frac{1}{h^2}\log^2\frac{Rh}{\delta} + \mathcal{J}_0(h)\log\frac{R}{\delta} + \mathcal{J}_{00}(h)\log^2\frac{R}{\delta} \right]. \\
	}
%	\eqs{
%	f_1^W(h,\delta) &\equiv \frac{2\omega(h)}{8\pi G}\int_{\delta/h}^{R}\diff\rho\ a_W\sqrt{\det g_W} \notag\\
%	&= \frac{L^2}{4\pi G}\left[ -\frac{1}{h^2} + \tilde{\mathcal{J}}^W_{2}(h)\frac{R^2}{\delta^2} + \tilde{\mathcal{J}}^W_{1}(h)\frac{R}{\delta} + \tilde{\mathcal{J}}^W_{0}(h) \right]\log\frac{L}{\alpha\delta}, \label{eq:gW1}\\
%	f_1^J(h,\delta) &\equiv \frac{2\omega(h)}{8\pi G}\int_{\delta/h}^{R}\diff\rho\ a_{J_\pm}\sqrt{\det g_{J_\pm}} \notag\\
%	&= \frac{L^2}{8\pi G}\Bigg[\frac{1}{2h^2}\left(\log^2\frac{R^2}{K\delta^2} - \log^2Kh^2\right) + \mathcal{J}^J_{00}(h)\log^2\frac{R}{\delta} + \mathcal{J}^J_{0}\log\frac{R}{\delta}  \notag\\
%	&\qquad + \left(-\frac{2}{h^2}\log\frac{Rh}{\delta} + \tilde{\mathcal{J}}^J_{0}(h)\log\frac{R}{\delta} + \tilde{\mathcal{J}}^J(h) \right)\log\frac{2 L^2}{\alpha\beta\delta^2} \Bigg], \\
%	f_1^E(h,\delta) &\equiv \frac{2\omega(h)}{8\pi G}\int_{\delta/h}^{R}\diff\rho\ a_E\sqrt{\det g_E} \notag\\
%	&= \frac{L^2}{4\pi G}\Bigg[\frac{1}{h^2}\left(\log^2Kh^2 - \log^2\frac{R}{Kh\delta}\right) + \mathcal{J}^E_{00}(h)\log^2\frac{R}{\delta} + \mathcal{J}^E_0\log\frac{R}{\delta} \notag\\
%	&\qquad + \left(\frac{2}{h^2}\log\frac{Rh}{\delta} + \tilde{\mathcal{J}}^E_{0}(h)\log\frac{R}{\delta} + \tilde{\mathcal{J}}^E(h) \right)\log\frac{L}{\beta\delta} \Bigg].
%	}
We thus extract the divergence coefficients
\eq{
	& F^{(1)}_{\rm joint} = \frac{L^2R}{4\pi G} \times 2 ,\\
	& F^{(0)}_{\rm joint} = -\frac{L^2}{4\pi G} \left[ -\frac{2+2\log(h_0)}{h_0} + \int_0^{h_0}\diff h\ \mathcal{J}_0(h) \right] ,\\
	& F^{(00)}_{\rm joint} = \frac{L^2}{4\pi G} \left[ -\frac{1}{h_0} + \int_0^{h_0}\diff h\ \mathcal{J}_{00}(h) \right].
}
%	where the $h=\delta/R$ contributions give
%	\eqs{
%	&F_h^W - \int_0^{\delta/R}\diff h\ \tilde{\mathcal{J}}^W_{2}(h)\frac{R^2}{\delta^2} = \frac{L^2}{4\pi G}\left(\frac{1}{h_0} - \frac{2R}{\delta}\right)\log\frac{L}{\alpha\delta}, \label{eq:gW2}\\
%	&F_h^J = \frac{L^2}{8\pi G}\Bigg[ \left( 4\log\frac{R}{\delta}-2\log K-4\right)\frac{R}{\delta} + \frac{2\log K}{h_0}\log\frac{R}{\delta} - \frac{2}{h_0}\log^2\frac{R}{\delta} \notag\\
%	&\qquad + \left(- \frac{2R}{\delta} + \frac{2}{h_0}\log\frac{R}{\delta} + \frac{2}{h_0}(1+\log h_0)\right)\log\frac{2L^2}{\alpha\beta\delta^2} \Bigg], \label{eq:FhJ}\\
%	&F_h^E = \frac{L^2}{4\pi G}\Bigg[ \left(-4\log\frac{R}{\delta}+2\log K+6\right)\frac{R}{\delta} - \frac{2}{h_0}(1+\log Kh_0)\log\frac{R}{\delta} + \frac{1}{h_0}\log^2\frac{R}{\delta} \notag\\
%	&\qquad + \left( \frac{2R}{\delta} - \frac{2}{h_0}\log\frac{R}{\delta} - \frac{2}{h_0}(1+\log h_0)\right)\log\frac{L}{\beta\delta} \Bigg]. \label{eq:FhE}
%	}
%	Note that the second lines in eq(\ref{eq:FhJ}) and eq(\ref{eq:FhE}) are exactly opposite, indicating no dependence on $\beta$. Roughly speaking, these contributions are from $\partial A$. It is interesting to see that although the $\beta$ dependence does not vanish in general, it might only come from the interior of the boundary region $A$. 
with integrands as follows:
\eq{\label{eq:Js}
	& \mathcal{J}_{0 }(h) = \frac{2}{h^2}\left[-\log(h) + \frac{H_+H_-}{Kh^2(1+h^2)}\log\frac{H_\pm}{h^2(1+h^2)}\omega(h) + \frac{1+h^2}{Kh^2}\log(h)\omega(h) \right], \\
	& \mathcal{J}_{00}(h) = \frac{1}{h^2}\left[-1 + \left(\frac{H_+H_-}{Kh^2(1+h^2)}-\frac{Kh^2}{1+h^2}\right)\omega(h) \right].
}
Then we turn to the null hypersurface contributions $I_c$ as in \eqref{eq:hyper}. The volumes of $W$, $J$ and $E$ are directly given by integrations of the volume forms, derived in Appendix~\ref{ap:geo}. As shown in \eqref{eq:Ic}, this contribution can be rearranged into a volume term, an entropy term and an additional sub-leading term proportional to $V_E-V_J$. As the first two are trivial or well studied, we only focus on the third term, which evaluates as
\eq{\label{eq:ellp}
	\ell'(h_0) \equiv \frac{1}{L^2}(V_E - V_J) &= \frac{1}{L^2}\int_{A_\delta} (\sqrt{\det g_E} - \sqrt{\det g_J}) \\
	&= \log\frac{R}{\delta}\int_0^{h_0}\diff h\ \frac{2K}{1+h^2}\omega(h).
}
Note that $V_E$ and $V_J$ individually has area law divergences, but they cancel when taking the difference, so this additional piece of contribution does not contain new area law terms.

%%%%%%%%%%%%%%%%%%%%%%%%%%%%%%%%%%%%%%%%%%%%%%%%%%%%%%%%%%%%%%%%%

\subsection{Counter term contributions}
\label{sec:ct_computation}

Finally, let's look at the counter term contributions, and also focus on the $L$ independent part. The $L$ dependent part can be combined with the joint contributions, and is already discussed at the end of last section. 

Starting from \eqref{eq:ct_calc}, we first perform the $\lambda$ integration. For hypersurface $\mathcal{E}$ we have
\eq{\label{eq:ct_beta}
	& \int_0^{\lambda_*}\diff\lambda\sqrt{\det g(\lambda)}\ \Theta(\lambda)\log\frac{\delta|\Theta(\lambda)|}{2} \\
	= & -\frac{L^2(1+h^2)}{K\rho h^4}\Bigg[ \frac{K^2h^2(\rho h - \delta)^2}{2(\rho(1+h^2) \pm Kh\delta)^2} \left(1+2\log\frac{Kh^2(\rho h-\delta)(\rho+\rho h^2\pm Kh\delta)}{\delta H_\pm(\rho H_\mp \pm 2Kh\delta)}\right) \\
	& \qquad \qquad + \log\left(1-\frac{K^2h^2(\rho h-\delta)^2}{(\rho+\rho h^2\pm Kh\delta)^2}\right) \Bigg]
}
and for $\mathcal{W}$ we have
\eq{\label{eq:ct_alpha}
	& \int_0^{\tilde{\lambda}_*}\diff{\tilde\lambda}\sqrt{\det \tilde{g}(\tilde\lambda)}\ \Theta(\tilde\lambda)\log\frac{\delta|\Theta(\tilde\lambda)|}{2} \\
	= & -\frac{L^2(1+h^2)(\rho H_\mp \pm 2Kh\delta)}{2K h^4} \\
	& \times \left[ \frac{h^2}{H_+\delta^2} - \frac{H_\pm}{(\rho+\rho h^2 \pm Kh\delta)^2}\left(1 + 2\log\frac{h(\rho+\rho h^2 \pm Kh\delta)}{\delta H_\pm}\right) \right].
}
The upper signs are for convex kinks and the lower signs for concave kinks.

The $\Delta I_{\mathcal{W}}$ contains quadratic and linear divergence:
\eq{
	& F^{(2)}_{\mathcal{W}} = -\frac{1}{2}F^{(2)}_{\rm bulk} ,\\
	& F^{(1)}_{\mathcal{W}} = -\frac{1}{4}F^{(1)}_{\rm bulk} - \frac{L^2R}{4\pi G}\times 2.
}
The $\log$ divergence coefficients are
\eq{
	& F^{(0)}_{\mathcal{E}} = -\frac{L^2}{8\pi G} \int_0^{h_0}\diff h\ \mathfrak{d}_0^{\mathcal{E}}(h) ,\\
	& F^{(0)}_{\mathcal{W}} = -\frac{L^2}{8\pi G} \left[\frac{3+2\log(h_0)}{h_0} + \int_0^{h_0}\diff h\ \mathfrak{d}_0^{\mathcal{W}}(h) \right].
}
with integrands
\eq{
	& \mathfrak{d}^{\mathcal{E}}_{0}(h) = \left[\frac{K}{1+h^2}\left(1+2\log\frac{Kh^3}{1+h^2}\right) + \frac{H_+H_-}{Kh^4(1+h^2)}\log\frac{2H_+H_-}{(1+h^2)^2}\right]\omega(h), \\
	& \mathfrak{d}^{\mathcal{W}}_{0}(h) = \frac{1}{h^2}\left[ 1+2\log h -\frac{H_+H_-}{Kh^2(1+h^2)}\left(1-2\log\frac{H_\pm}{h(1+h^2)}\right) \omega(h)\right].
}
There are also $\log^2$ divergence from counter terms
\eq{ 
	& F^{(00)}_{\mathcal{E}} = \frac{L^2}{8\pi G}\int_0^{h_0}\diff h\ \frac{K}{1+h^2}\omega(h) ,\\
	& F^{(00)}_{\mathcal{W}} = \frac{L^2}{8\pi G} \left[\frac{1}{h_0} + \int_0^{h_0}\diff h\ \frac{1}{h^2}  \left(1-\frac{H_+H_-}{Kh^2(1+h^2)}\omega(h) \right)\right].
}

%%%%%%%%%%%%%%%%%%%%%%%%%%%%%%%%%%%%%%%%%%%%%%%%%%%%%%%%%%%%	
\subsection{Total results}

Now we can wrap up all the contributions. We find some exact cancellations:
\eq{
	& F^{(2)}_{\rm total} = F^{(2)}_{\rm bulk} + 2F^{(2)}_{\mathcal{W}} = 0 ,\\
	& F^{(00)}_{\rm total} = F^{(00)}_{\rm joint} + 2F^{(00)}_{\mathcal{E}} + 2F^{(00)}_{\mathcal{W}} = 0,
}
which indicates the absence of $\log^2$ divergence, and the quadratic divergence only comes from the volume term in \eqref{eq:Ic}. The linear divergence does not vanish after summing over all contributions
\eq{\label{eq:def_Delta}
	F^{(1)}_{\rm total} &= F^{(1)}_{\rm bulk} + F^{(1)}_{\rm joint} + 2F^{(1)}_{\mathcal{W}} \\
	&= \frac{L^2R}{4\pi G}\left[1 + \frac{1}{2}\int_0^{h_0}\diff h\ \mathcal{B}_1(h) \right] \\
	&\equiv \frac{L^2R}{4\pi G}\Delta(h_0).
}
We discuss about the function $\Delta(h_0)$ in Sec~\ref{sec:geo}. Thus the final action reads

\eq{\label{eq:I_total}
		I_{\rm total} & = \frac{V_{\delta}}{4\pi G}\log\frac{c}{\alpha} - \frac{1}{\pi}S_{\rm EE}\log\frac{2c}{\alpha} \\
		& + \frac{L^2}{4\pi G}\left[ \Delta(h_0)\frac{R}{\delta} + \left(\ell(h_0) + \ell'(h_0)\log\frac{2c^2}{\alpha\beta} \right)\log\frac{R}{\delta} \right] + \mathcal{O}(\delta^0),
	}
where the first line comes from \eqref{eq:Ic}, $\ell'(h_0)$ is derived in \eqref{eq:ellp}, and $\ell(h_0)$ is given by
\eq{\label{eq:ell}
	\ell(h_0) &= - \left( F^{(0)}_{\rm bulk} + F^{(0)}_{\rm joint} + 2F^{(0)}_{\mathcal{E}} + 2F^{(0)}_{\mathcal{W}} \right)/\frac{L^2}{4\pi G} \\
	&= -\frac{2}{h_0} + \int_0^{h_0}\diff h\ \left(\mathcal{B}_0(h) + \mathcal{J}_0(h) + \mathfrak{d}^{\mathcal{W}}_0(h) + \mathfrak{d}^{\mathcal{E}}_0(h) \right) \\
	&= -\frac{2}{h_0} - \int_0^{h_0}\diff h\ \frac{2}{h^2}\Bigg[1 - \frac{1+h^2}{Kh^2}\left(1+\log\frac{1+2h^2+h^4-K^2h^4}{(1+h^2)^2}\right)\omega(h) \\
	& \quad - \frac{Kh^2}{1+h^2}\left(1-\log\frac{1+2h^2+h^4-K^2h^4}{Kh^2(1+h^2)}\right)\omega(h) \mp \frac{3(1+h^2)^2-K^2h^4}{2(1+h^2)H_\pm}\omega(h) \Bigg].
	}

%--------------------- App B ------------------------------------------------------------------------
	
	\section{Induced geometry on  null hypersurfaces}
	\label{ap:geo}
	
	In this appendix, some detailed calculations about the null hypersurface geometry will be carried out. 
	% These results are involved in several discussions like Sec(\ref{sec:issues}) and the joint contribution computations. 
	
	Because we are always using $w^\alpha = (\rho,\theta)$ defined on the HRT surface as coordinates, the light sheet geometry is mostly convenient to be studied as induced geometry following the flow of $V_{\pm}^{\mu}$ and $U_{\pm}^{\mu}$. Starting from the HRT surface, from eq(\ref{eq:para_E}) we get the induced metric on $\mathcal{E}^{\pm}$:
	\eq{
		& g_{\alpha\beta}(\lambda) = \frac{\partial X^{\pm}_\mu(\lambda)}{\partial w^\alpha}\frac{\partial X^{\pm}_\nu(\lambda)}{\partial w^\beta}g^0_{\mu\nu}, \\
		& \sqrt{\det g(\lambda)} = \frac{(1+h^2)(\lambda_c^2-\lambda^2)\rho}{Kh^2L^2}.
	}
	As $\lambda$ is chosen to be affine parameter, the expansion rate of the congruence $V_{\pm}^{\mu}$ can be computed directly
	\eq{\label{eq:Theta_E}
		\Theta(\lambda) = \frac{1}{\sqrt{\det g(\lambda)}}\frac{\partial}{\partial\lambda}\sqrt{\det g(\lambda)} = -\frac{2\lambda}{\lambda_c^2-\lambda^2}.
	}
	As expected, the monotonicity is predicted by focusing theorem, and it blows up to negative infinity at the caustic $\lambda = \lambda_c$. The flow ends at $\lambda = \lambda_* < \lambda_c$ given by \eqref{eq:lambdaE_star}.

	Next we investigate the hypersurfaces $\mathcal{W}^{\pm}$. With eq(\ref{eq:para_W}), we write down the induced metric
	\eq{
		& \tilde{g}_{\alpha\beta}(\eta) = \frac{\partial \tilde{X}^{\pm}_\mu(\eta)}{\partial w^\alpha}\frac{\partial \tilde{X}^{\pm}_\nu(\eta)}{\partial w^\beta}g^0_{\mu\nu}, \\
		& \sqrt{\det \tilde{g}(\eta(\tilde{\lambda}))} = \frac{(1+h^2)(H_\pm L^2+\tilde{\lambda} h(\rho+\rho h^2\pm Kh\delta))^2(\rho H_\mp \pm 2Kh\delta)}{Kh^4H_\pm L^2(\rho+\rho h^2\pm Kh\delta)^2}.
	}
	Here we reparameterize with affine parameter $\tilde{\lambda}$ for convenience
	\eq{
		\eta(\tilde{\lambda}) = \frac{\tilde\lambda h^2(\rho(1+h^2) \pm Kh\delta)^2}{H_\pm[H_\pm L^2 + \tilde\lambda h(\rho(1+h^2) \pm Kh\delta)]}.
	}
	and we drop the $\pm$ superscript due to time reflection symmetry. We also restore the $>$ and $<$ superscripts for concave and convex kinks respectively. The end value of $\tilde\lambda$ on the surface $W$ is given by the condition
	\eq{
		\tilde{X}^z(\tilde\lambda_*) = \delta,
	}
	which is solved by
	\eq{\label{eq:lambdaW_star}
	\tilde\lambda^{\stackrel{<}{>}}_* = \frac{(\rho h - \delta)\sqrt{1+h^2+h^{\prime2}}}{\delta(\pm\delta + \rho h\sqrt{1+h^2+h^{\prime2}})}
	}
	We also obtain the expansion rate on $\mathcal{W}$ as
	\eq{\label{eq:Theta_W}
		\Theta(\tilde\lambda) = -\frac{2h(\rho(1+h^2) \pm Kh\delta)}{H_\pm L^2 + \tilde{\lambda}h(\rho(1+h^2) \pm Kh\delta)}.
	}

	Finally, we recognize the induced metric on the joint surfaces as $g$ and $\tilde{g}$ at different ends of the affine parameter ranges, and obtain the relations:
	\eq{
		& \sqrt{\det g(0)} \equiv \sqrt{\det g_E} = \frac{(1+h^2)}{Kh^4\rho}L^2, \\
		& \sqrt{\det g^{\stackrel{<}{>}}(\lambda_*)} = \sqrt{\det \tilde{g}^{\stackrel{<}{>}}(0)} \equiv \sqrt{\det g^{\stackrel{<}{>}}_J} = \frac{(1+h^2)H_\pm(\rho H_\mp \pm 2Kh\delta)}{Kh^4(\rho(1+h^2)\pm Kh\delta)^2}L^2, \\
		& \sqrt{\det \tilde{g}^{\stackrel{<}{>}}(\eta_*)} \equiv \sqrt{\det g^{\stackrel{<}{>}}_W} = \frac{(1+h^2)(\rho H_\mp \pm 2Kh\delta)}{Kh^2H_\pm\delta^2}L^2.
	}
% These are the area elements used in the joint contributions. 

%--------------------- The End ------------------------------------------------------------------------

\bibliography{complexity}
\bibliographystyle{JHEP}
\end{document}